\documentclass[fleqn,usenatbib]{mnras}

\usepackage{newtxtext,newtxmath}

\usepackage[T1]{fontenc}
\usepackage{ae,aecompl}

\usepackage[pdfpagelabels=false]{hyperref}	
\hypersetup{colorlinks=true,linkcolor=blue,citecolor=blue,filecolor=blue,urlcolor=blue,}

\usepackage{graphicx}	
\usepackage{amsmath}	
\usepackage{amssymb}	

\newcommand{\halomod}{\textsc{halomod}}
\newcommand{\HI}{H\textsc{i}}
\newcommand{\mmin}{\ensuremath{M_{\rm min}}}
\newcommand{\mmax}{\ensuremath{M_{\rm max}}}

\title{Intensity mapping cross-correlations II: HI halo models including shot noise}

\author[L.Wolz et al.]{
L. Wolz,$^{1,2}$\thanks{E-mail: lwolz@unimelb.edu.au}
S. G. Murray,$^{3,2}$
C. Blake$^{4,2}$
and J.S. Wyithe$^{1,2}$
\\
$^{1}$School of Physics, University of Melbourne, Parkville, VIC 3010, Australia\\
$^{2}$ARC Centre of Excellence for All-Sky Astrophysics (CAASTRO)\\
$^3$ICRAR, Curtin Institute of Radio Astronomy, GPO Box U1987, Perth, WA 6845, Australia\\
$^4$Centre for Astrophysics \& Supercomputing, Swinburne University of Technology, P.O. Box 218, Hawthorn, VIC 3122, Australia\\
}

\date{Accepted XXX. Received YYY; in original form ZZZ}

\pubyear{2017}

\begin{document}
\label{firstpage}
\pagerange{\pageref{firstpage}--\pageref{lastpage}}
\maketitle

\begin{abstract}
\HI\ intensity mapping data traces the large-scale structure matter distribution using the integrated emission of neutral hydrogen gas (\HI ). The cross-correlation of the intensity maps with optical galaxy surveys can mitigate foreground and systematic effects, but has been shown to significantly depend on galaxy evolution parameters of the \HI\ and the optical sample. Previously, we have shown that the shot noise of the cross-correlation scales with the \HI\ content of the optical samples, such that the shot noise estimation infers the average \HI\ masses of these samples. In this article, we present an adaptive framework for the cross-correlation of \HI\ intensity maps with galaxy samples using our implementation of the halo model formalism (Murray et al 2018, in prep) which utilises the halo occupation distribution of galaxies to predict their power spectra. We compare two \HI\ population models, tracing the spatial halo and the galaxy distribution respectively, and present their auto- and cross-power spectra with an associated galaxy sample. We find that the choice of the \HI\ model and the distribution of the \HI\ within the galaxy sample have minor significance for the shape of the auto- and cross-correlations, but highly impact the measured shot noise amplitude of the estimators, a finding we confirm with simulations. We demonstrate parameter estimation of the \HI\ halo occupation models and advocate this framework for the interpretation of future experimental data, with the prospect of determining the \HI\ masses of optical galaxy samples via the cross-correlation shot noise.
\end{abstract}




\section{Introduction}
The cosmological evolution of our Universe can be tested via probes of the statistics of large-scale structure. Common techniques include measuring the Baryon Acoustic Oscillations (BAOs), which act as a standard ruler for distance measures constraining the Cosmic acceleration (see e.g. \citealt{2012MNRAS.426.2719R,2014MNRAS.441...24A}), as well as galaxy clustering which employs the positions of galaxies to measure their cosmological power spectrum (for instance \citealt{2005PhRvD..71j3515S,2007ApJ...657..645P}). Both BAO measurements and galaxy clustering require the determination of millions of galaxy positions over large volumes in order to minimize statistical uncertainties. Traditionally, optical telescopes have been employed for cosmological measurements as radio telescopes are limited in sensitivity. Beyond the nearby Universe, the determination of redshifts at radio frequencies, in which the spectrum is close to featureless, is extremely challenging. The most notable radio spectral line, at a rest-wavelength of 21cm, is caused by the spin-flip of the neutral hydrogen (\HI) and is comparably weak. It has only been directly detected up to $z=0.36$ in a single object \citep{2016AAS...22732304F}, and in the statistically averaged spectrum via \HI\ stacking up to $z\approx 0.32$ \citep{2018MNRAS.473.1879R}. To circumvent these limitations, \HI\ intensity mapping provides a novel technique to map the large-scale structure distribution as traced by neutral hydrogen gas via low-resolution observations of the integrated and unresolved 21cm emission of multiple objects.

After \HI\ intensity mapping was proposed as a test of cosmology more than a decade ago (see \citealt{2004MNRAS.355.1339B,2008MNRAS.383.1195W,2008PhRvL.100i1303C}), \cite{2009MNRAS.394L...6P} reported the first detection of structure in \HI\ intensity maps of the local Universe. Later  \cite{2010Natur.466..463C} reported a detection in observations around $z\approx 0.8$. The challenges in detecting the \HI\ intensity mapping power spectrum arise due to the weakness of the redshifted \HI\ signal in comparison to the radio foregrounds in combination with the radio telescope systematics, see e.g. \cite{2015ApJ...815...51S,2017MNRAS.464.4938W,2017arXiv171107843H}. The cross-correlation signal of an \HI\ map with an overlapping galaxy survey is insensitive to many of these systematics and increases the significance of detection. The detection of the cosmological distribution via the power spectrum \citep{2013ApJ...763L..20M} was achieved by measurements of the Green Bank telescope at medium redshift $z=0.8$ in cross-correlation with the WiggleZ Dark Energy survey \citep{2010MNRAS.401.1429D}, constraining the \HI\ energy density and the \HI\ bias to $\Omega_{\rm HI} b_{\rm HI}=0.63 \substack{+0.23 \\ - 0.15} \times 10^{-3} $ \citep{2013MNRAS.434L..46S}. A more recent detection has also been made, using the cross-correlation of the \HI\ intensity maps of the Parkes telescope with the 2dF Galaxy Redshift Survey at $z\approx 0.08$ \citep{2017arXiv171000424A}. The analysis presents a 5-sigma detection of the cross-power spectrum with a significant drop of the power on smaller scales, $k \approx 1.5 h /{\rm Mpc}$, indicating a strong anti-correlation of \HI\ with the red galaxy sample.

The future of \HI\ intensity mapping looks very promising as a large number of purpose-built instruments are under design and construction. The instruments can be divided into 3 categories: single dish telescopes similar to the pioneering Green Bank and Parkes telescopes as well as equipped with multi-beam receivers (e.g. BINGO \citealt{2013MNRAS.434.1239B}), dish interferometers such as HIRAX \citep{2016SPIE.9906E..5XN}, and cylindrical dish interferometers such as CHIME \citep{2014SPIE.9145E..22B} or Tianlai \citep{2012IJMPS..12..256C}. Additionally, the Square Kilometre Array, an international radio interferometer with unprecedented scale and sensitivity, will conduct \HI\ intensity mapping for wide ranges of redshifts $0<z<6$ \citep{2015aska.confE..19S}. Two SKA pathfinder projects, MeerKAT and the Australian SKA Pathfinder (ASKAP), are capable of intensity mapping observations, and will be able to explore different observational techniques such as the employment of the array in single-dish mode \citep{2017arXiv170906099S} or phased array feeds, in preparation for the SKA observations to commence in the next decade. Forecasts predict that the future SKA \HI\ intensity mapping experiments will be able to measure distances via BAOs to a level that is comparable to  Stage IV optical experiments as well as obtaining new constraints on higher, unobserved redshifts \citep{2015ApJ...803...21B}. The forth coming intensity maps will also set new constraints on non-Gaussianity through measuring the ultra-large scales of the power spectrum \citep{2014JPhCS.566a2004C}. For all mentioned experiments, the cross-correlation of the \HI\ intensity mapping signal with galaxy surveys will be a crucial test for systematics, and most likely be the first observable to deliver new scientific results.

In addition to cosmological parameters, the amplitude and the clustering power of the \HI\ intensity mapping power spectrum depends on the distribution of the neutral hydrogen gas with respect to the underlying matter field, and additionally for cross-correlations on the observed optical galaxy sample. Recently, \HI\ models based on available data (see e.g. \citealt{2017MNRAS.469.2323P,2017arXiv171201296P}) as well as predictive theories \citep{2012IJMPS..12..256C} have been proposed to deliver the theoretical framework for interpretation of the intensity mapping signal. 

In this work we extend existing \HI\ models to predict the cross-correlation of intensity maps with galaxy surveys to enhance the interpretation of existing and forth coming data and provide a framework to include halo occupation parameters into the cosmological analysis of future measurements (for forecasts see \citealt{2015PhRvD..92j3506P,2016JApA...37...26S,2017MNRAS.470.4251P}). In \cite{2017MNRAS.470.3220W} we have shown that the shot noise in the cross-power spectrum, which is caused by the discrete nature of galaxy data, scales with the average \HI\ mass per optical galaxy. Hence, intensity mapping data can be employed to determine an average \HI\ mass for any over lapping galaxy sample. This allows determination of global scaling relations between star-formation activity as traced by the optical sample and their gas contents, for redshifts well beyond the current limits for direct gas detection. In this work, we present a theoretical framework which correctly determines the shot noise contribution given the \HI\ parameters of the distribution, and which can be employed to fit the \HI\ parameters and shot noise in future observational data.

In this paper, we first briefly introduce the halo model framework, along with our chosen numerical implementation (\halomod) in Sec.~\ref{Sec-hm}. Here we also introduce the employed halo occupation models for galaxies and \HI\ models and present theoretical euqtaions for the \HI\ auto-power spectra and their respective cross-power spectra. In Sec.~\ref{Sec-LN} we describe our method of producing lognormal realisations of joint optical and \HI\ samples, which we will use to verify our theoretical formalism. In Sec.~\ref{Sec-Poisson}, we review the current understanding of shot noise on power spectra and discuss its implementation in \halomod. We present and examine the comparison of theory with lognormal simulations in Sec.~\ref{Sec-comp}. In the following Sec.~\ref{Sec-MCMC}, we demonstrate how \halomod\ can be used to constrain \HI\ parameters via MCMC parameter estimation. We discuss our findings and present the conclusions in Sec.~\ref{Sec-sum}.

\section{Halo model description}
\label{Sec-hm}
The halo model \citep{2000MNRAS.318.1144P,2002PhR...372....1C} is a highly successful description of the cosmological density field that uses empirical models of the internal properties of dark matter halos to access non-linear scales.
It has been employed, along with a prescription for the abundance of galaxy tracers within halos termed the \textit{halo occupation distribution} (HOD), to predict the spatial statistics of various galaxy populations, typically in order to constrain various properties of the selected sample \citep{2005ApJ...633..791Z,2011ApJ...736...59Z,2013MNRAS.429.3604B}.
It has recently been extended to the domain of \HI\ abundance by \citet{2017MNRAS.464.4008P,2017MNRAS.469.2323P}.

The essence of the halo model consists of the assumption that all material is sequestered into discrete halos, which are in turn self-similar objects that scale exclusively as a function of their mass. 
Consequently, knowledge of the spatial arrangement of the halo centres combined with a knowledge of their internal profiles, how these scale with the halo's mass, and the abundance of halos at any given mass, yields a full statistical description of the matter field down to arbitrarily small scales in real space.
Likewise, assuming that any given tracer inhabits halos with an abundance exclusively as a function of their mass, the statistics of the tracer field may also be determined.

In summary, to describe the two-point statistics of a tracer field (or the cross-correlation of tracers), one requires the following ingredients:
\begin{enumerate}
	\item The non-linear matter power spectrum \citep{2003MNRAS.341.1311S}.
    \item The radial profile of the tracer within the halo, $\rho(r)$; we typically employ the standard NFW profile \citep{1997ApJ...490..493N}, but also check the modified, or ``cored'' NFW employed by \cite{2017MNRAS.464.4008P}.
    \item The mass function of halos, $n(m)$; we use the fitting formula of \cite{2008ApJ...688..709T}.
    \item The abundance and distribution of tracers within halos, $N(m)$; we describe our choices for this component further in \S\ref{sec:hod}.
    \item The concentration-mass relation, $c(m)$, which defines how the profile scales with halo mass; we use the fit of \citet{2008MNRAS.390L..64D}.
    \item The bias of halos of a given mass, $b(m)$; we use the function determined by \citet{2010ApJ...724..878T}.
\end{enumerate}
Additionally, the effects of halo exclusion can be modeled, such that pairs of the tracer that are very close are probabilistically assigned to the same halo and excluded from the counts between different halos, to avoid double-counting. We omit this modeling for this introductory work, but note that its inclusion is trivial within the \textsc{halomod} package that we use.

\subsection{The \textsc{halomod} package}
All halo model calculations performed in this work use the \textsc{halomod} \textsc{Python} library\footnote{Source code at \url{https://github.com/steven-murray/halomod}.} (Murray et al., 2018, \textit{in prep.}). 
This library is built on the \textsc{hmf} package\footnote{Available at \url{https://github.com/steven-murray/hmf}} \citep{2013A&C.....3...23M}, which handles the cosmology, linear power spectra, and mass functions. 
The \textsc{halomod} code provides many models for halo profiles, halo bias, concentration-mass relations, HODs and halo exclusion, along with the necessary framework to combine these to produce spatial statistics.

A key feature of the \textsc{hmf} framework which is extended to \textsc{halomod} is the simplicity of defining new component models and ``plugging'' them into the calculations. 
Thus for instance it is simple to define a new HOD model from a standard galaxy HOD, and is instantly usable within the framework without having to modify the source code. We note that versions of both \textsc{hmf} and \textsc{halomod} that calculate the results of this paper can be obtained via the {\tt feature/HIHOD} branch of each. 
\subsection{Halo occupation distributions}
\label{sec:hod}
In our study we require several HOD models: one which describes the full galaxy count population, another which describes a particular optically-selected sample count, and a model which describes the \HI\ occupation. We use variants of the simple HOD parameterisation of \cite{2005ApJ...630....1Z} (Z05) in all cases. This model depends on three parameters: the minimum halo mass to be occupied by a galaxy $\mmin$,  the characteristic halo mass $M_1$ which marks the turn-over of the broken power-law, and the power-law coefficient $\alpha$ of the satellite HOD. We extend the parametrisation by adding the maximum (cut-off) halo mass $\mmax$ as a parameter.

In general the HOD can be split into two separate classes of objects;  central galaxies located at the centre of the halo, and satellite galaxies that trace the halo's density profile. The Z05 model assigns the following parameterisations to each component:
\begin{align}
\langle N_{\rm cen}(m) \rangle &=
  \begin{cases}
    1\phantom{(m/M_1)} &  \mmin < m < \mmax \\
    0 &  \text{otherwise}
  \end{cases} \\
\langle N_{\rm sat}(m) \rangle &=
  \begin{cases}
    (m/M_1)^\alpha & \mmin < m < \mmax \\
    0 & \text{otherwise}
  \end{cases}
\end{align}

When stating that a sample may be described by a separation of central and satellite galaxies, we furthermore assume (in this paper) that this separation is due to the central having a much higher probability of existence within the sample than its associated satellites. This may be understood easily in terms of optical samples, in which the central galaxy is typically much brighter than the satellites. To approximate the effect of this a priori knowledge, our \halomod\ algorithms assert in such cases that a central galaxy \textit{must} be present before any satellites. In this case, the average total occupation is accurately given by the following definition,
which ensures that the total occupation is zero whenever the central occupation is zero, but otherwise yields the expected sum of central and satellite:
\begin{equation}
	\langle N(m)\rangle = \langle N_{\rm cen}(m)\rangle(1 + \langle N_{\rm sat}(m)\rangle).
\end{equation}

\halomod\ does not limit the form or parametrisation of the HODs and more complex models can be assumed. 

In this article, we adopt two fiducial galaxy HODs, in this study, referred to as \textit{sample} and \textit{field}, where we assume that the galaxy \textit{field} model is a description of all optically-detectable and \HI\ emitting galaxies and \textit{sample} is an optically-detected sub-sample of the \textit{field}. The HOD parameters of all galaxy and \HI\ models can be found in Tab.~\ref{tab:model_params}, where we choose representative values for all parameters in our toy models.  

In the following, we postulate two variations of the \HI\ HOD model in the framework of the halo model. For demonstration purposes, we base the parametrisation of the \HI\ HODs on Z05. More physically-motivated and data-driven models, such as \cite{2017MNRAS.464.4008P} and \cite{2017arXiv171204469P}, can be easily implemented in \halomod\ and studied with our methods.

\paragraph*{ Continuous HI distribution. }
In this scenario we assume that the \HI\ continuously traces the dark matter halo following an independent density profile, for example a cored NFW profile as in \cite{2017MNRAS.469.2323P}.
This implies that the \HI\ is not associated with galaxies and there are no central or satellite contributions to the density. This model is best suited to describe the cold gas distribution at the early stages of galaxy formation at the end of the epoch of reionisation and resembles semi-numerical approaches to cosmological simulations of intensity maps (e.g. \citealt{2014MNRAS.444.3183A}). 

We alter the Z05 HOD to describe the \HI\ mass distribution, by adding an extra normalisation $A_{\rm HI}$ in units of $M_\odot/h$ to scale the distribution to produce typical \HI\ masses, increasing the number of parameters to five. 
\begin{equation}
\langle M_{\rm HI}(m) \rangle=\begin{cases}
    A_{\rm HI}~((m/M_1)^\alpha + 1 )& \mmin < m < \mmax \\
    0 & \text{otherwise}
  \end{cases}
  \label{equ-HODHIcont}
\end{equation}
The additional term $+1$ in the HOD is introduced to simplify comparison with the our second \HI\ model.

\paragraph*{Discrete HI distribution. }
In this model, we assume that the \HI\ is \textit{on average} following the underlying dark matter halo density profile throughout the halo, but specify that the \HI\ mass is co-located with the underlying galaxy \textit{field}. Thus the \HI\ in any given halo is discretely located. This model describes a stage of galaxy evolution in which most \HI\ is confined within galaxies and inter-galactic cold gas is negligible in the intensity maps. The approach predicts a similar distribution to semi-analytic simulations which model the cold gas abundances within star-forming regions (e.g. \citealt{2014MNRAS.440..920L,2017MNRAS.465..111K}).

We model this case in a similar fashion to galaxies, in which we split the HOD contributions into central and satellite components. \begin{equation}
\langle M_{\rm HI}^{\rm cen}(m)\rangle=\begin{cases}
    A_{\rm HI} & \mmin < m < \mmax \\
    0 & \text{otherwise}
  \end{cases}
\end{equation}
and the satellite part by 
\begin{equation}
\langle M_{\rm HI}^{\rm sat}(m) \rangle=\begin{cases}
    A_{\rm HI}~\left[(m/M_1)^\alpha \ \right] & \mmin < m < \mmax\\
    0 & \text{otherwise}
  \end{cases}
\end{equation}

We note that this \HI\ model has a dependence on the model defining the underlying galaxy \textit{field} with which it is co-located. While the HOD itself as defined above requires no knowledge of the underlying \textit{field} HOD, and thus the auto-power spectrum of the \HI\ is fully self-defined, its cross-correlation with  an optical sample relies on the actual distribution of \HI\ within the halo, which we have described as being dependent on the \textit{field}. We will see that this information will be necessary in order to define theoretical cross-correlations, Poisson noise, and also to self-consistently produce joint simulations.

\begin{table}
    \centering
    \caption{HOD parameters for all models considered in this work. All masses are given as $\log_{10}$ and in units of $M_\odot/h$.}
    \label{tab:model_params}
    \begin{tabular}{ cccccccc}
        \hline
		Model & $M_{\rm  min}$ & $M_{\rm max}$ & $\alpha$ & $M_{1}$ & $\log{A_{\rm HI}}$ \\ 
		\hline
		Galaxy \textit{field} &11.0 & 17.0 & 0.5 & 11.0 & -  \\
		Galaxy \textit{sample} & 11.5 & 17.0 & 0.45 &11.0 & -   \\
		\HI\ continuous &11.0 & 17.0 & 0.7 & 11.0 & 11.0 \\
		\HI\ discrete & 11.0 & 17.0 & 0.7 & 11.0 & 11.0 \\
	\end{tabular}
\end{table}

\subsection{Galaxy Power Spectra}
The power spectrum is divided into its 2-halo and 1-halo contribution, $P(k)=P_{2h}(k)+P_{1h}(k)$. The 2-halo term closely follows the linear matter power spectrum $P_{\rm lin}(k)$ and, in its most general form applicable to cross-correlation on large scales, the 2-halo term is expressed as 
\begin{equation}
P_{2h}^{ij}(k)=b_{i}(k) b_{j}(k) * P_{\rm lin}(k),
\label{EquP2halo}
\end{equation}
where $b_{i}$ is the effective bias of the $i$th probe, given as 
\begin{equation}
b_i(k)=\frac{1}{\bar{n}_g} \int {\rm d}m \ n(m) \ b(m) \langle N^i(m)\rangle \ u_i(k|m).
\label{Equbias}
\end{equation}
Here $b(m)$ is the halo bias, $u(k|m)$ is the Fourier transform of the halo mass profile following the NFW model, and $\bar{n}_g$ is given by the number density of the galaxies, computed as 
\begin{equation}
\bar{n}_g = \int {\rm d}m \ n(m)  \langle N(m)\rangle.
\end{equation}
$n(m)$ is the halo mass function, for more details on the implementation please refer to \cite{2013A&C.....3...23M}.

The 1-halo term is given by the clustering within the halos and depends on the number of centrals and satellite galaxies. 
For the auto-correlation of one probe, this results in
\begin{equation}
	\begin{aligned}
	P_{1h}(k) =\frac{1}{\bar{n}_g^2} \int {\rm d}m \ n(m) \left[ \vphantom{\frac{1}{2}} \right. & \langle N_{\rm cen}N_{\rm sat}\rangle   \ u(k|m) \\ + 		& \left. \frac{1}{2}\langle N_{\rm sat}(N_{\rm sat}-1)\rangle \ u^2(k|m) \right].
	\end{aligned}
	\label{EquPgal0}
\end{equation} 
The first term depends on the expectation value of the number of central-satellite pairs per halo multiplied by the halo mass profile and the second term on the expectation value of the number of satellite-satellite pairs per halo mass multiplied by the self-convolved mass profile. 
Since in our model there can only ever be either zero or one central galaxy in a halo, and under the assumption that the central galaxy is always the first of the halo to be included in a sample, we have $\langle N_{\rm cen} N_{\rm sat}\rangle = \langle N_{\rm cen} \rangle \langle N_{\rm sat} \rangle  $. 
Furthermore, for Poisson-distributed $X$, $\langle X(X-1) \rangle \equiv \langle X \rangle^2$, which means (assuming the satellite occupation is Poisson-distributed) that\begin{equation}
	P_{1h}(k) =\frac{1}{\bar{n}_g^2} \int {\rm d}m \ n(m) \left[ \langle N_{\rm cen} \rangle \langle N_{\rm sat}\rangle   \ u(k|m) + \frac{1}{2}\langle N_{\rm sat}\rangle^2 \ u^2(k|m) \right],
		\label{EquPgal}
\end{equation}
This form is convenient, as it only depends on the mean occupation functions which we have defined above.

For the cross-correlation of two different galaxy samples which follow different HODs and density profiles, the analogue of Eq.~\ref{EquPgal0} is
\begin{equation}
\begin{aligned}
P_{1h}^{ij}(k) =\frac{1}{\bar{n}_i \bar{n}_j} \int {\rm d}m \ n(m) \left[\vphantom{N_{\rm cen}^j} \right. &  \langle N^i_{\rm cen}N^j_{\rm sat}\rangle  \ u_j(k|m) + \\
&\langle N^j_{\rm cen}N^i_{\rm sat}\rangle  \ u_i(k|m) +\\
& \left. \langle N^i_{\rm sat}N^j_{\rm sat}\rangle \ u_i(k|m)\ u_j(k|m) \ \right].
\label{EquPX1halo}
\end{aligned}
\end{equation}

In general we cannot further reduce this equation, because it is not guaranteed that the absence of a central galaxy in one sample necessitates the absence of satellites (as well as central) in a different sample. However, if the central HOD happens to be a step-function, so that at any mass either all or none of the haloes have centrals, the central-satellite term decomposes as before. We note that this is an \textit{extra condition}, which was not required for Eq.~\ref{EquPgal}.
This allows us to re-write the equation as follows:
\begin{equation}
	\begin{aligned}
	P_{1h}^{ij}(k) =\frac{1}{\bar{n}_i \bar{n}_j} & \int {\rm d}m  \ n(m) \left[\vphantom{N_{\rm cen}^j} \right. \langle N^i_{\rm cen}\rangle \langle N^j_{\rm sat}\rangle  \ u_j(k|m) + \\
	& \langle N^j_{\rm cen}\rangle \langle N^i_{\rm sat}\rangle \ u_i(k|m) +\\
	& \left. \left(\langle N^i_{\rm sat}\rangle \langle N^j_{\rm sat}\rangle + \sigma_i \sigma_j R^{ij} - Q \right) \ u_i(k|m)\ u_j(k|m) \ \right],
\end{aligned}
\label{EquPXcorr}
\end{equation}
where $R^{ij}$ is the correlation of the satellite occupation between the probes, and $\sigma_i$ the standard deviation of the satellite occupation, which for a Poisson occupation is simply $\sqrt{\langle N_{\rm sat}\rangle}$. $Q$ is equal to the expected number of shared points between the samples unless either tracer is continuously spatially distributed which results in $Q=0$. 

In general, $R^{ij}$ is constrained to be within $(-1,1)$ and depends on the complicated physical interactions of the two tracer populations. However, for the toy models we employ in this paper, it is possible to provide a better description which we present in detail in Appendix~\ref{Appcorrcont} and \ref{Appcorrdisc}.

\subsection{HI Power Spectra}
Following the same arguments as the previous section, we may derive the power spectrum of \HI\ density fluctuations for both cases presented in Sec.~\ref{sec:hod}. The 2-halo term of the \HI\ power spectra for both models is similar to Eqs.~\ref{EquP2halo} and \ref{Equbias} with the galaxy HOD substituted by the \HI\ occupation $\langle M_{\rm HI}(m)\rangle$ of the respective model, such that
\begin{equation}
b_{\rm HI}(k)=C_{\rm HI} \int {\rm d}m \ n(m) \ b(m) \langle M_{\rm HI}(m)\rangle \ u_{\rm HI}(k|m).
\label{EqubiasHI}
\end{equation}
where the coefficient $C_{\rm HI}$ is described below.
The \HI\ halo density profile $u_{\rm HI}(k|m)$ is commonly defined as a modified (or cored) NFW profile \citep{2017MNRAS.469.2323P} which in real space reads as
\begin{equation}
\rho_{\rm HI}(r) = \frac{\rho_0 r_s^3}{(r+0.75r_s)(r+r_s)^2}
\end{equation}
where $r_s$ is the scale radius of the dark matter halo which is defined as $r_s=r_{\rm vir}/c(m)$ and $r_{\rm vir}$ is the virial radius of the halo. We refrain from the use of a \HI\ specific parametrisation of $r_s$ and adopt the concentration-mass relation fit from \citet{2008MNRAS.390L..64D}. In \halomod\ we employ the analytic expression of the Fourier transform $u_{\rm HI}(k)$ of this profile \citep{2017MNRAS.469.2323P}.

\HI\ intensity maps are measured in brightness temperature $T_{\rm HI}$. To follow this convention, we convert all power spectra into temperature units, using a conversion $C_{\rm HI}$, given by
\begin{equation}
C_{\rm HI}=\frac{3A_{12}h_{\rm P}c^3 (1+z)^2}{32\pi m_{\rm H}k_{\rm B}\nu_{\rm 21}^2 H( z)}
\end{equation}
with $h_{\rm P}$ the Planck constant, $k_{\rm B}$ the Boltzmann constant, $m_{\rm H}$ the mass of
the hydrogen atom, $A_{12}$ the emission coefficient of the 21cm line transmission and $\nu_{\rm 21}$ the rest frequency of the 21cm emission. $H( z)$ is the Hubble parameter at redshift $z$. All presented studies are for redshift $z\approx0$. The plotted \HI\ power spectra are given in units of $\rm{ K^2 (Mpc}/h)^3$ and cross-power spectra as ${\rm K (Mpc}/h)^{3}$ if not stated otherwise.

The predicted mean brightness temperature for each \HI\ model can be determined via 
\begin{equation}
\overline{T_{\rm HI}}=C_{\rm HI} \int {\rm d}m \ n(m) \langle M_{\rm HI}(m)\rangle
\label{EquTmean}
\end{equation}
The mean \HI\ brightness temperature is directly proportional to the \HI\ energy density $\Omega_{\rm HI}$ which makes it a desired observable when conducting \HI\ intensity mapping experiments.

\paragraph*{Continuous HI distribution. }
The 1-halo term of the auto-power spectrum in this case, with lack of satellite components, can be written as 
\begin{equation}
	P^{\rm HI, cont}_{1h}(k)=  C_{\rm HI}^2 \int {\rm d}m \ n(m) \langle M_{\rm HI}(m)\rangle ^2  \ u_{\rm HI}(k|m)^2 ,
	\label{EquPHIcont}
\end{equation}
while the cross-correlation with a galaxy sample $g$ is
\begin{equation}
	\begin{aligned}
	P^{g\rm HI, cont}_{1h}(k)= \frac{ C_{\rm HI}}{\bar{n}_g} & \int {\rm d}m \ n(m)  u_{\rm HI}(k|m) \\ 
	& \times \left[ (u_g(k|m) \ \langle N^g_{\rm sat}(m)\rangle  \langle M_{\rm HI}(m)\rangle  + R^{g\rm HI} ) \right. \\
	& \left. + \ \langle N^g_{\rm cen}(m)\rangle  \langle M_{\rm HI}(m)\rangle \right]
	\end{aligned}
\end{equation}
where $R^{g\rm HI}$ is a galaxy-\HI\ correlation coefficient.
As there is no central-satellite split in the \HI\ HOD, the clustering is simplified into two terms - one in which the satellite galaxies pair with the  \HI\ profile, and another in which the single (possible) central galaxy pairs with the \HI\ profile. We fiducially consider a value of  $ R = 0$ for this work, which implies that the \HI\ mass is uncorrelated with the galaxy occupation. The more detailed derivation of the correlation factor $R$ and an example for a correlated toy model can be found in Appendix~\ref{Appcorrcont}.

\paragraph*{Discrete HI distribution. } 
The 1-halo power spectrum of the discrete \HI\ model can be written similarly to Eq.~\ref{EquPgal}, assuming that the positions of the satellite occupation are Poisson-distributed: 
\begin{equation}
	\begin{aligned}
	P^{\rm HI,dsc}_{1h}(k)=C_{\rm HI}^2 \int {\rm d}m \ n(m) \left[ \vphantom{\frac{1}{2}}\right. & 		u_{\rm HI}(k|m) \langle M_{\rm HI}^{\rm sat}(m) \rangle  \langle M_{\rm HI}^{\rm cen}(m)  \rangle\\
    & +\ \left. \frac{1}{2} \langle M_{\rm HI}^{\rm sat}(m) \rangle ^2 u_{\rm HI}(k|m) ^2 \right].
	\end{aligned}
\end{equation}
The 1-halo term of the \HI\ cross-correlation with a galaxy sample reads as
\begin{equation}
	\begin{aligned}
	P^{g\rm HI,dsc}_{1h}(k) =\frac{ C_{\rm HI}}{\bar{n}_g} \int {\rm d}m & \ n(m)  \left[\vphantom{N_{\rm cen}^j} \right.   \left(\langle N^g_{\rm cen}\rangle  \langle M_{\rm HI}^{\rm sat} \rangle \right) \ u_{\rm HI}(k|m) + \\
	&\left(\langle  M_{\rm HI}^{\rm cen} \rangle \langle N^g_{\rm sat}\rangle \right) \ u_g(k|m) +\\
	& \left. \left(\langle N^g_{\rm sat}\rangle \langle M_{\rm HI}^{\rm sat} \rangle \right) \ u_g(k|m)\ u_{\rm HI}(k|m) \ \right].
\end{aligned}
\end{equation}
We note the absence of the correlation term, $R$. This is due to exact co-location of the \HI\ with the optical galaxies, as explained in detail in Appendix~\ref{Appcorrdisc}. Briefly, in this model, \HI\ abundance depends only on the properties of the galaxy in which it is situated, and this galaxy, by construction, has no correlation with other galaxies. Therefore all correlations are expressed at a separation of zero, and do not affect the shape of the 1-halo term. This may alternatively be seen as the exact cancellation of the correlation term with the Q term in Eq.~\ref{EquPXcorr}.

\begin{figure}
\includegraphics[width=0.5\textwidth]{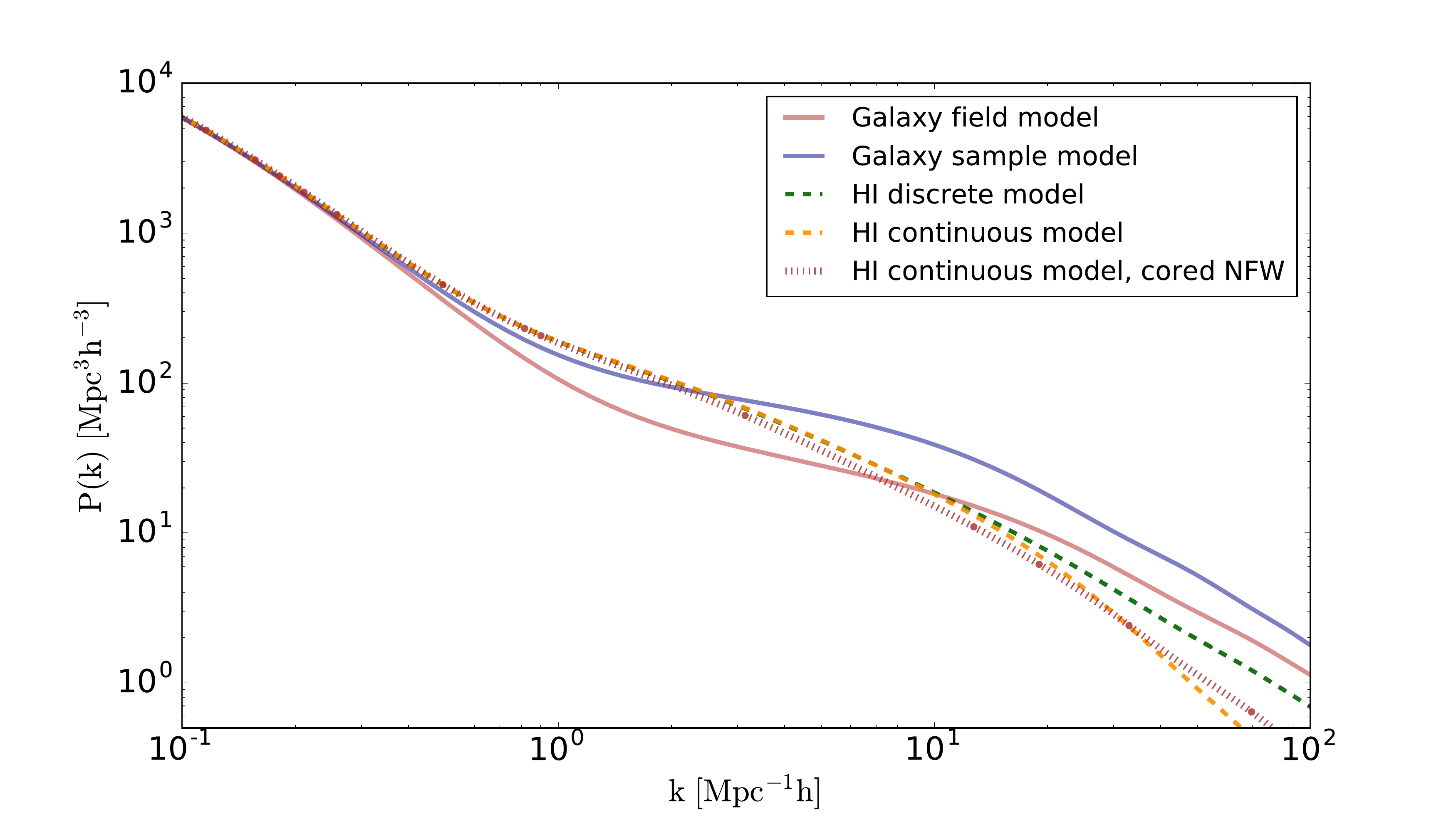}
\caption{The auto-power spectra predicted by our model for the case of galaxy \textit{field} population, galaxy \textit{sample}, \HI\ continuum, and \HI\ discrete model. The \HI\ power spectra are normalised by the square of the mean temperature predicted by each model using Eq.~\ref{EquTmean} for presentation purposes. Note that by construction, both \HI\ models predict the same mean brightness temperature.}
\label{PSautohalomod}
\end{figure}

\begin{figure}
\includegraphics[width=0.5\textwidth]{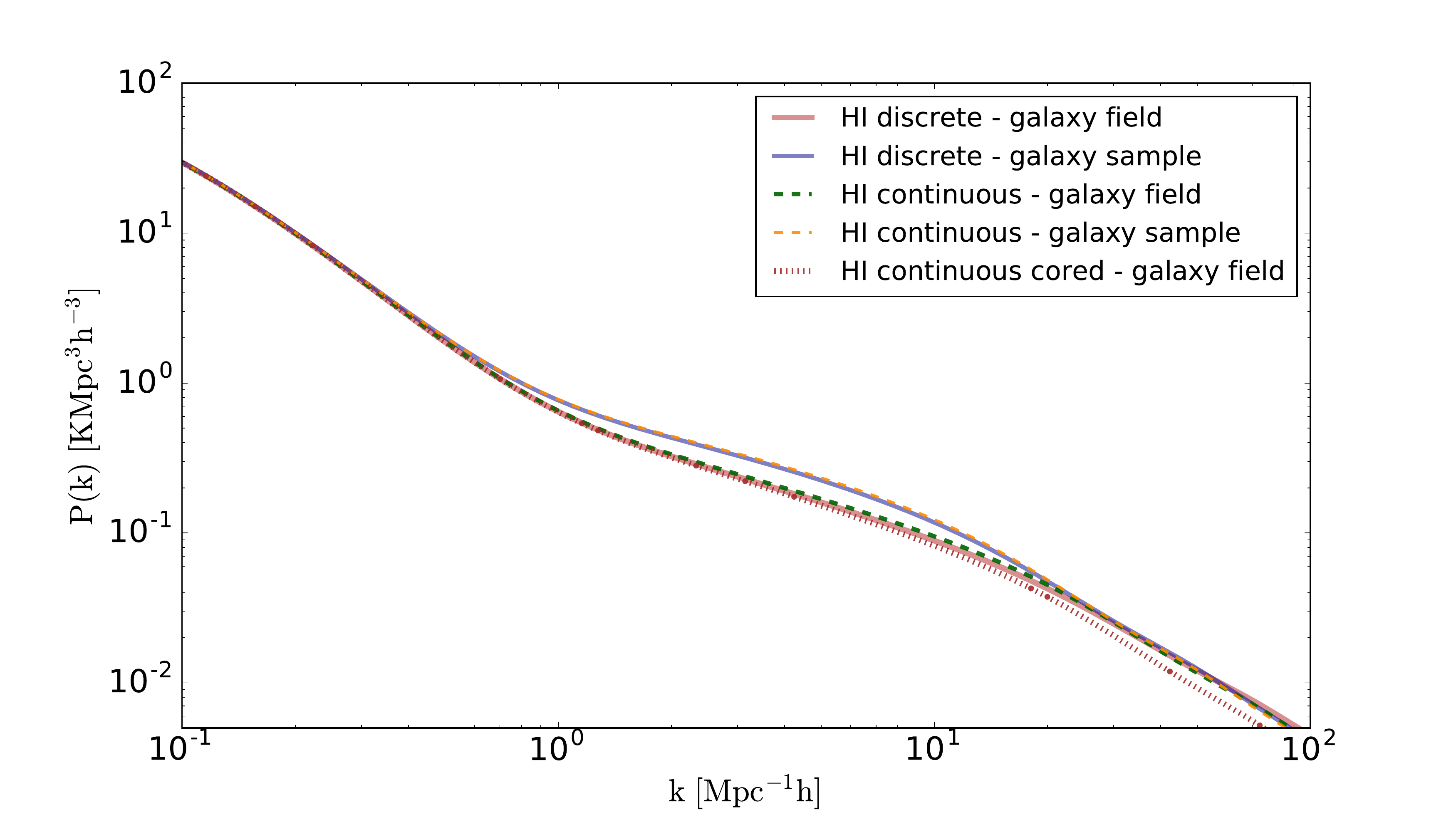}
\caption{The cross-power spectra predicted by our model for the case of galaxy field population, galaxy sample, \HI\ continuum, and \HI\ discrete model. The \HI\ cross-power spectra are normalised by the mean temperature predicted by each model using Eq.~\ref{EquTmean} for presentation purposes. Note that by construction, both \HI\ models predict the same mean brightness temperature.}
\label{PScrosshalomod}
\end{figure}

\paragraph*{}The auto-power spectrum predictions of \halomod\ are shown in Fig.~\ref{PSautohalomod}, where we show the two models of galaxy power spectra, called \textit{field} and \textit{sample}, with HOD parameters defined in Tab.~\ref{tab:model_params}. In yellow and green, we compare the \HI\ power spectra of the continuous and discrete models, where we renormalise the \HI\ spectra through division by $\overline{T_{\rm HI}}^2$ . We can see that for $k < 10h/{\rm Mpc}$ both models' predictions closely agree. For $k>10h/{\rm Mpc}$, the continuous \HI\ model falls off more quickly since the 1-halo term contains no central-satellite contribution  in the discrete case. 
In this figure, we also demonstrate how the cored \HI\ profile alters the power of the 1-halo term in comparison to the standard NFW profile. In the remainder of the article, we employ the standard NFW profile for our computations such that the comparison of the cases are focused on their clustering terms rather than the impact of the density profile.

The cross-power spectrum prediction of both \HI\ models with the two galaxy models are shown in Fig.~\ref{PScrosshalomod}. The differences in the two \HI\ models are negligible over all scales $k$. Furthermore, even the differences in the two different galaxy models are very small compared to the variation in their auto-power spectrum. The agreement of the two models is by construction as they follow the same \HI\ HOD parameters and we implement the continuous case to be the sum of the central and satellite terms of the discrete model. Therefore, they correlate in a similar fashion with the galaxy samples which is not to be expected in a general case.

In both figures, we neglect the shot noise, also referred to as the Poisson Noise (PN) contribution $P_{\rm PN}$, and we will discuss its contribution in detail in the following sections.
\section{Lognormal simulations}
\label{Sec-LN}
In order to test the accuracy of the analytic routines within \halomod, we create a number of mock realisations of the tracer populations. 
As this is done explicitly to test the routines, the simulations are prepared to mimic the assumptions of the halo model formalism at the simplest level.
In this section we describe the method used to generate these simulations.
\subsection{Galaxy populations}
We consider a cube of volume $L^3 ({\rm Mpc}/h)^{3}$ with $N^3$ grid cells, in which we generate a lognormal density field \citep{1991MNRAS.248....1C} using the \textsc{powerbox} package\footnote{Available at \url{https://github.com/steven-murray/powerbox}.}. We choose matching input power spectra and parameters to \halomod\ to ensure comparability of the results using the Planck15 cosmological model \citep{2016A&A...594A..13P}.

We choose a minimum halo mass $ M_{\rm min, h}$ such that all halos containing galaxies in our sample lie above the threshold.
We then draw a number density of halo masses $n_h = \int_{M_{\rm min, h}} n(m) {\rm d}m$ from the halo mass function distribution. 
These halos are placed probabilistically within the grid volume, with the probability of landing in a certain cell given by its relative density.
The final positions of each halo are drawn randomly within each cell, rendering sub-grid scales highly inaccurate.
We note that the mass of each halo does not affect its placement, which effectively means that the halo bias is unity for all masses when comparing simulations to theory, we therefore set the theoretical halo bias to unity in \halomod.

Finally, we use the resultant halo catalog, with masses and positions, as the scaffolding on which to assign the tracer population. 
Here we will describe the methods used for producing a single tracer population, suitable for comparing to auto-spectra.
We use a routine in which for each halo $i$ we perform the following steps:
\begin{enumerate}
 \item Sample a single number (zero or one) $C_i$ from a Bernoulli distribution with mean $\langle N_{\rm cen}(m_i)\rangle$
    \item If $C_i=1$, place a galaxy at $\vec{x}_i$ and continue, else proceed to next halo.
    \item Sample a number $N_{\rm sat}^i$ from a Poisson distribution with mean $\langle N_s(m_i) \rangle$.
    \item If $N_s^i > 0$, sample $N_{\rm sat}^i$ radii, $r^i_j$ from the halo's profile, $\rho(r,m)$, and sample $(\theta_j, \phi_j)$ isotropically to yield 3D co-ordinates, $\vec{x}^i_j$ centred at the origin.
    \item Assign $N_{\rm sat}^i$ galaxies to positions $\vec{x}_i + \vec{x}^i_j$. 
\end{enumerate}
We note that this procedure does not take into account halo exclusion -- halos are allowed to overlap arbitrarily -- and thus to reproduce the results analytically also requires no halo exclusion model.

In our simulations, we first apply the steps outlined above using the \textit{field} HOD to create a galaxy catalogue which is assumed to contain all available galaxies. 
We then create a sub-sample of the galaxy catalogue which follows the HOD of the galaxy \textit{sample}. Similarly to steps $(i)$ and $(ii)$, for each galaxy in the \textit{field} catalogue we draw a single number $C_i \in (0,1)$ from the Bernoulli distribution with mean $P_i=\langle N_{\rm sample}(m_i) \rangle/\langle N_{\rm field}(m_i) \rangle$ to determine if the galaxy is part of the \textit{sample}. The positions of the galaxies are kept identical. We note that this procedure does not strictly retain the Poisson-distributed nature of the satellite galaxies in the \textit{sample}. Nevertheless, the mean is retained, and we do not expect the departure from Poisson to be significant.
\label{Secgalpop}
\subsection{HI populations}
\paragraph*{Continuous HI distribution. }
For the continuous model, we assign an \HI\ mass to each halo produced by the lognormal realisations, where we draw the \HI\ masses according to the input \HI\ HOD at halo mass $m$ assuming a Gaussian distribution with a standard deviation $\sigma_{\rm HI}=0.25\langle M_{ \rm HI}(m)\rangle$. In order to mimic the continuous \HI\ distribution throughout the halo, we convolve the resulting \HI\ mass with a density profile using the method as follows. We note that any arbitrary density profile independent of the underlying halo density profile can be used in this routine. 

According to the convolution theorem, the convolution of the \HI\ masses with any given profile is a multiplication in Fourier space which is more computationally efficient. However, generally the halo profile is a function of halo mass $m_i$. In order to reduce computation, we apply a projection algorithm for the convolution. The \HI\ masses are therefore binned according to their halo mass into $N_{\rm bin}$ bins. We create $N_{\rm bin}$ cubes with each \HI\ mass located at their respective halo centre position. Each cube is Fourier transformed and multiplied by the Fourier-transformed profile of the mean halo mass $\overline m_i$ of the respective halo mass bin. We then sum all cubes to create the final intensity mapping cube. For the case of the NFW profile, the algorithm converges for $N_{\rm bin}=25$. 

We note that the continuous \HI\ distribution is based on the same underlying halo distribution of the lognormal realisation but is independent of the \textit{field} or \textit{sample} galaxy densities and satellite positions.

\paragraph*{Discrete HI distribution. } 
In the discrete model, the \HI\ HOD is associated with an underlying galaxy \textit{field} HOD $\langle N_{\rm field}\rangle$ which describes the distribution of all \HI\ emitting objects. In our algorithm, the galaxy \textit{field} is drawn from $\langle N_{\rm field}\rangle$ as described in Sec.~\ref{Secgalpop}. We then assign the \HI\ mass of each galaxy from a Gaussian distribution with mean $\langle M_{ \rm HI, field}\rangle=\langle M_{\rm HI}(m_i)\rangle / \langle N_{\rm field}(m_i)\rangle$ and standard deviation $\sigma_{\rm HI}=0.25\langle M_{ \rm HI, field}(m_i)\rangle$ for satellites and centrals respectively. The assumption that $\langle M_{ \rm HI, field}\rangle=\langle M_{\rm HI}(m_i)\rangle / \langle N_{\rm field}(m_i)\rangle$ is only true if the probability of selecting a galaxy is independent of \HI\ mass, which precludes the use of this  algorithm for creating correlated samples. This model allows for the galaxy \textit{field} HOD and the \HI\ HOD to follow independent models and parametrisations within the limitation that $M_{\rm min}$ and $M_{\rm max}$ of the \HI\ sample cannot be outside the defined galaxy mass range. In our study, we choose $M_{\rm min, HI}=M_{\rm min, field}$ and $M_{\rm max, HI}=M_{\rm max, field}$ for simplicity. Additionally, the \HI\ mass can be scaled by an independent \HI\ density profile, similarly to the continuous case. In our study, we set the \HI\ profile equal to the NFW profile of the underlying galaxy \textit{field}.
\subsection{Correlated populations}
\label{SecHIpopcorr}
The procedures described above produce galaxy catalogues and \HI\ intensity maps useful for determining their 1-halo clustering and Poisson noise. It is non-trivial to populate a physically-motivated model for correlated galaxy-\HI\ samples in the framework of \halomod . Commonly, the \HI\ mass of galaxies is associated with their star-formation activity and other more complex mechanisms depending on the galaxy's evolutionary state, which is beyond the scope of our work.  

As previously stated, in the continuous \HI\ case the correlation factor $R$ is determined through the dependence of the galaxy numbers on the \HI\ mass per halo, or vice versa ( a basic example of this is set out in Appendix~\ref{Appcorrcont}), and this impacts the 1-halo contribution of the cross-power spectrum. 

For the discrete \HI\ case, we demonstrated that a correlation between the \HI\ distribution and the galaxy abundances has no impact on the 1-halo term.
However, if \HI\ masses and the galaxy abundances are correlated, the averaged \HI\ mass per galaxy over the \textit{sample} is modulated and hence the amplitude of the cross-shot noise is changed, as we detail in the following section. As one of our primary concerns is investigating the cross-shot noise, we demonstrate this effect with correlated simulations through the following procedure.

In order to create a \HI\ correlation, we either up- or down-weight the \HI\ masses of the galaxies in the \textit{sample} by drawing for each galaxy $i$ a Gaussian variable $\delta M_{i,\rm HI}$ with zero mean and multiplying the absolute value $|\delta M_{i,\rm HI}|$ with a weighting factor $w=\{+1, -1\}$. When assigning \HI\ masses, we then add $w \times |\delta M_{i,\rm HI}|$ to the mean \HI\ mass at $m$ given by the \HI\ HOD $\langle M_{ \rm HI}(m)\rangle$. This implies that all galaxies in the \textit{sample} either have higher or lower \HI\ mass than the mean of the Gaussian. Galaxies which are not part of the \textit{sample} are not affected by the weighting and their \HI\ masses fluctuate around the mean. This process slightly alters the measured brightness temperature of the \HI\ intensity maps. However, if the galaxy \textit{sample} is a small enough sub-sample of the whole galaxy \textit{field}, this effect will be minor. 
\section{Poisson Noise}
\label{Sec-Poisson}
\subsection{Auto power spectra }
The additive shot noise contribution to the power spectrum, also referred to as Poisson noise in the literature, is due to the finite number of data points used to probe a continuous field. In galaxy surveys, the shot noise is caused by the finite number of galaxies in the sample employed to trace the matter field. The resulting Poisson noise on the power spectrum is scale-independent with the amplitude equal to the inverse of the galaxy density:
\begin{equation}
P_{\rm PN}=\frac{1}{\overline{n_{g}}}=\frac{1}{N_{\rm g}/V}.
\end{equation}
The total power measured from galaxy survey data is $P(k)=P_{2h}(k)+P_{1h}(k)+P_{\rm PN}$.

The shot noise of a galaxy distribution is not strictly Poissonian. The deviations from the Poisson limit were examined by e.g. \citep{2011PhRvD..84h3509H,2013PhRvD..88h3507B,2017MNRAS.470.2566P}. The deviations are caused by halo exclusion, non-linear clustering on small scales and satellite galaxy distributions, where the fraction of satellite galaxies can determine if the noise is sub- or super-Poissonian \citep{2013PhRvD..88h3507B}.

In the halo model context, the shot noise of the halo power spectrum may be determined by the $k\rightarrow 0 $ limit of the 1-halo term of the power spectrum, which results in the Poisson limit. This approach is correct when treating tracers without sub-sampling the halo with satellite populations. For galaxy populations including a central / satellite split, the $k\rightarrow 0 $ limit of the 1-halo term does not result in the Poisson limit and over-estimates the shot noise. \cite{2017PhRvD..96h3528G} investigate the shot noise expression for dark matter, halos and tracers in the halo model framework, considering galaxy populations with satellites. They derive the correction terms to the 1-halo term to accurately determine the deviations from the Poissonian noise on scales $k\ll 1 h/{\rm Mpc}$. For our scales of interest where the shot noise dominates the overall power for $k\gg1 h/{\rm Mpc}$, the shot noise must converge towards the Poisson limit of the 1-halo term neglecting the satellite correlations \citep{2017PhRvD..96h3528G}. For the remainder of this study, we will only consider the Poisson limit of the shot noise and use the terms Poisson noise and shot noise interchangeably.

We derive the Poisson limit of the shot noise through 
\begin{equation}
P_{\rm PN}^{g}=\left( \int {\rm d}m \ n(m)\sum_{i=\rm sat, cen} \langle N^i(m) \rangle \right)^{-1}.
\label{EquPNgal}
\end{equation}

In intensity mapping, the nature of the shot noise depends on the \HI\ model used. In general, the shot noise is given by the standard deviation (or second moment) of the observed field (see \citealt{2017arXiv170909066K} and \citealt{2017MNRAS.467.2996B}), in this case the \HI\ mass distribution, such that
\begin{equation}
P_{\rm PN}^{\rm HI}=C_{\rm HI}^2 \int {\rm d}m \ n(m) \langle M_{\rm HI}(m)  \rangle^2.
\end{equation}
In the halo model context this is equal to taking the $k\rightarrow 0 $ limit of the 1-halo term neglecting the existence of satellite distributions, similar to in Eq.~\ref{EquPHIcont}, see also \cite{2017MNRAS.471.1788C} for a similar result.

In our specific case of the continuous model, the \HI\ masses are sampled per halo, which means that the number of samples is equivalent to the number of halos. However, the \HI\ is not discretely populated, but convolved with the halo profile which results in a continuous map of the \HI\ in voxel space. From a strict definition of Poisson noise originating from discrete sampling and resulting in a scale-independent noise, this means that the \HI\ continuous power spectrum does not contain a Poisson noise contribution. The absence of \HI\ shot noise in the continuous case is due to the strict smoothness of the \HI\ distribution tracing the halo. Alternatively, one could think of the 1-halo term as the Poisson contribution which is convolved by the halo profile. 

For our \HI\ discrete model, we assume that the \HI\ masses are sampled per galaxy, rather than halo, so we need to determine the second moment of the \HI\ distribution \textit{per galaxy} where the \HI\ per galaxy is given as $\langle M^i_{\rm HI,field}(m) \rangle = \langle M^i_{\rm HI}(m) \rangle / \langle N^i_{\rm field}(m) \rangle $ with $i=\{\rm cen, sat\}$, (again we note that this is strictly only correct if \HI\ and galaxy abundances are uncorrelated). The resulting Poisson noise of this model is
\begin{equation}
P_{\rm PN}^{\rm HI,dsc}= C_{\rm HI}^2 \int {\rm d}m \ n(m)\sum_{i=\rm sat, cen} \langle  M^i_{\rm HI,field}(m) \rangle^2 \langle N^i_{\rm field} \rangle .
\label{EquHISN}
\end{equation}
We do not usually know the HOD of the underlying galaxy \textit{field}, as well as the \HI\ HOD, in order to determine the \HI\ per galaxy as a function of halo mass. In practise, the Poisson noise can be modelled as a single additive number and fit to observations.

\subsection{Cross power spectra}
The shot noise in the cross-power spectrum of two galaxy samples, is determined by the galaxy density of the overlap of the two samples (see e.g. \citealt{2009MNRAS.400..851S}). If the two galaxy samples are mutually-exclusive, the amplitude of the Poisson noise in the power spectrum is zero. 

As outlined in the previous paragraph, the continuous \HI\ power spectrum does not contain a scale-independent Poisson noise contribution. Similarly, there is no Poisson noise generated in the cross-correlation of a continuum and a discrete galaxy sample, as the \HI\ distribution is assumed to be completely smooth and hence no additional sampling noise can correlate with the sampling noise of the galaxies. Again, alternatively, one could think of the sampling noise being incorporated in the 1-halo term as the Poisson contribution is convolved by the smooth \HI\ profile.

The discrete \HI\ model can be approached similarly to the case of two galaxy samples where shot noise is determined by the cross-section. In intensity mapping, it is assumed that each object emits \HI\ and contributes to the \HI\ maps. The cross-section of the \HI\ maps and the \textit{sample} is hence the number density of the \textit{sample} and the Poisson noise is inversely proportional to the galaxy number density. The \HI\ contribution to the Poisson noise is determined by the average \HI\ emission of the galaxies in the \textit{sample}. This general expression for the cross-shot noise can also be derived considering the $k\rightarrow 0 $ limit of the \HI-galaxy 1-halo term in absence of satellite populations. We derive the Poisson noise of the cross-correlation of the discrete case as
\begin{equation}
\begin{aligned}
P^{ g \rm HI,dsc}_{\rm PN}= & C_{\rm HI} \left( \int {\rm d}m \ n(m)  \sum_{i=\rm sat, cen} \langle  M^i_{\rm HI,field}(m) \rangle \langle N^i_{\rm sample} \rangle \right)  \\
&\times  \left( \int {\rm d}m n(m) \langle N(m) \rangle \right)^{-1}
\label{EqucrossSN}
\end{aligned}
\end{equation}
This equation agrees with the derivation in \cite{2017MNRAS.470.3220W}, where it was shown that the Poisson noise is directly proportional to the averaged \HI\ mass per galaxy in the \textit{sample}. This also implies that the amplitude of the Poisson noise is sensitive to any correlations between \HI\ and the abundance of galaxies in the \textit{sample}. In the following, we verify these expressions by comparing the \halomod\ predictions to simulations and showcase how the Poisson noise can be fit in order to determine the averaged  \HI\ masses of galaxy samples. 
\section{Comparison of \halomod\ with simulations}
\label{Sec-comp}
\subsection{Auto and cross power spectra}
We run a suite of lognormal simulations with different box sizes with length $L\in \{15, 25, 50,100\}{\rm Mpc}/h$ and $N=200$ pixels per side to create a valid comparison for all relevant scales $k$. We find that the \halomod\ prediction agrees well with the lognormal simulations on all scales. In order to resolve the scales dominated by the \HI\ and cross Poisson noise, we closely inspect simulations with volume $V=(15 {\rm Mpc}/h)^3$, which are presented in the following figures. We simulate 100 realisations of each lognormal field and the error bars of the following plots are given by the standard deviation of these realisations. 
 
\begin{figure}
\includegraphics[width=0.5\textwidth]{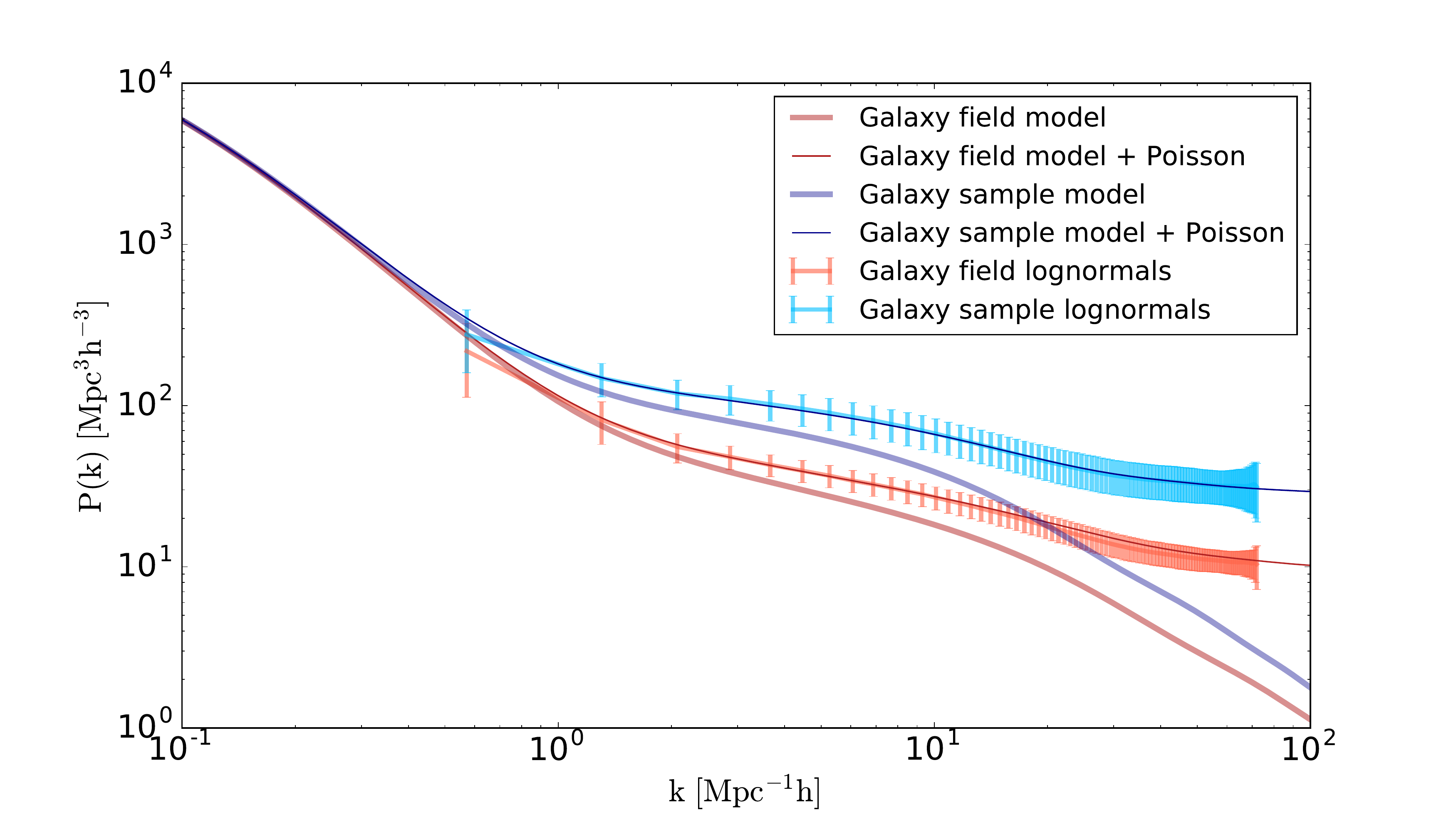}
\caption{The galaxy power spectra predicted by \halomod\ for the entire  galaxy field and the selected galaxy sample in comparison with an average power spectrum of 100 lognormal realisations with a box of length $15 {\rm Mpc}/h$ drawn from the respective galaxy HOD. We show \halomod\ predictions including and excluding Poisson noise contribution.}
\label{FigPgalcomp}
\end{figure}

In Fig.~\ref{FigPgalcomp}, we show the comparison of the auto-power spectra of the lognormal galaxy population models with the associated \halomod\  prediction. The power spectrum measurements from the lognormal realisations naturally contain Poisson noise contributions which we add to the \halomod\ predictions using Eq.~\ref{EquPNgal}. We see that the \halomod\ prediction including the Poisson noise is in agreement with estimates from the lognormal simulations. In this plot, we show our two galaxy models, \textit{field} and \textit{sample} (cf. Tab.~\ref{tab:model_params}). The galaxy densities of the populations are predicted by \halomod\ as $n_{\rm sample}=0.036 (h/\rm Mpc)^{3}$ and $n_{\rm field}=0.110\rm (h/\rm Mpc)^{3}$ and estimated from the realisations as  $n_{\rm sample}=0.036 \pm 0.0069\rm (h/\rm Mpc)^{3}$ and $n_{\rm field}=0.107\pm 0.017\rm (h/\rm Mpc)^{3}$.

\begin{figure}
\includegraphics[width=0.5\textwidth]{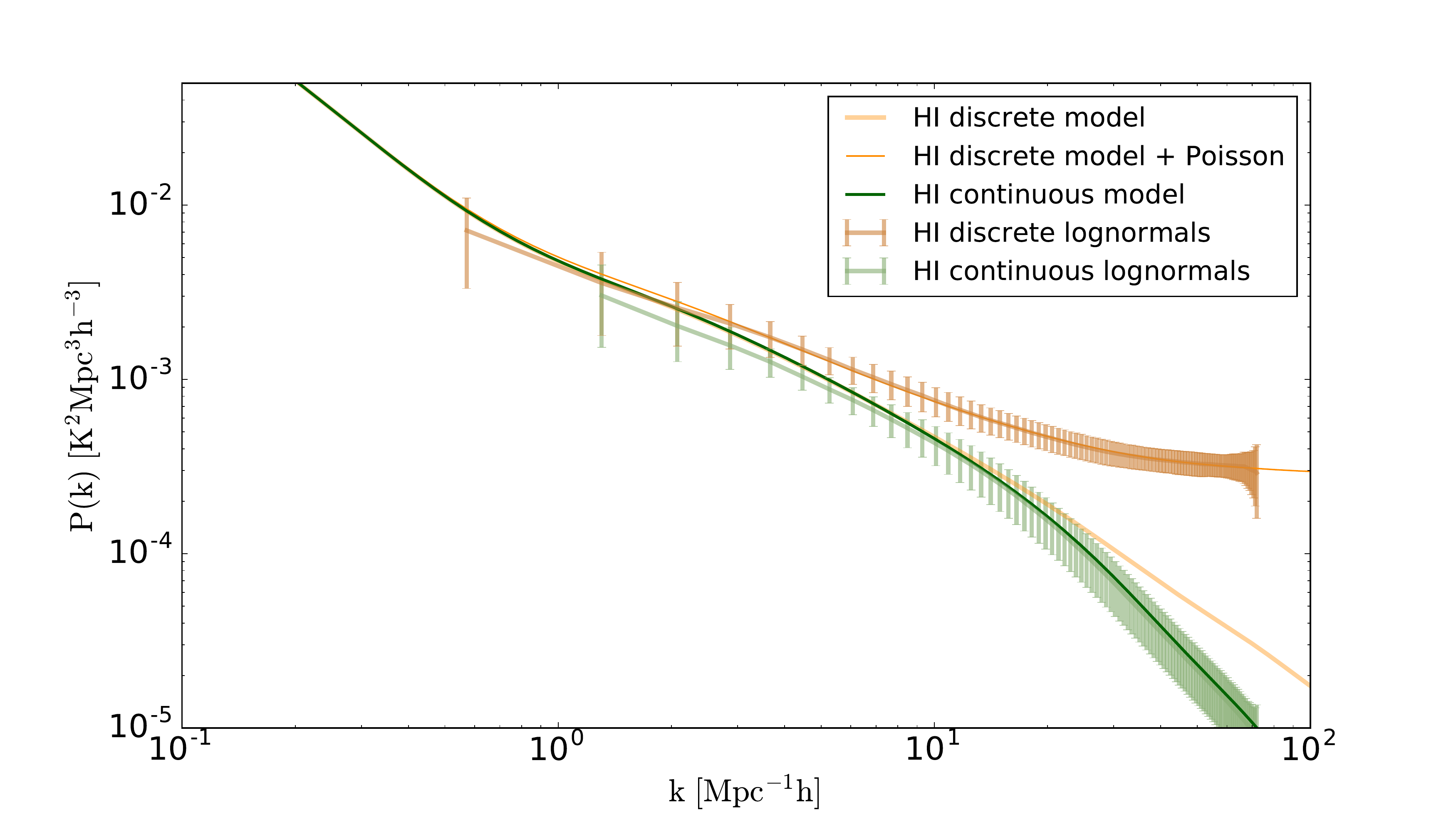}
\caption{The \HI\ power spectra predicted by \halomod\ for the \HI\ continuous model and the \HI\ discrete model in comparison with a average power spectrum of 100 lognormal realisations with a box of $15 {\rm Mpc}/h$ drawn from the respective galaxy HOD. Note that the \HI\ continuous model does not include a scale-independent Poisson noise contribution since it is estimated from a continuum field. We show \halomod\ predictions including and excluding Poisson noise contribution.}
\label{FigPHIcomp}
\end{figure}

In Fig.~\ref{FigPHIcomp}, the \HI\ power spectra of the lognormal realisations using the continuous and discrete model are shown in comparison to the analytic \halomod\ predictions in units of ${\rm K^2\ (Mpc}/h)^3$. For the continuous case describing smooth \HI\ distributions within halos independent of galaxy positions, we can see that the average of the simulations and the analytic prediction agree very well within the errors. In the discrete model, the \HI\ distribution is co-located with the galaxy positions and hence this model includes a Poisson noise contribution as described in Sec.~\ref{Sec-Poisson}. We add the theoretical prediction of the Poisson noise using Eq.~\ref{EquHISN} to the predictions of \halomod. The combined amplitude is in agreement with the estimates of the lognormal distributions. Both \HI\ models follow the same HOD parameter model, except the discrete model uses two additional parameters to describe the shape of the underlying galaxy field HOD ($\alpha_{\rm field}$ and $M_{1,\rm field}$). By construction, both \HI\ models predict the same \HI\ brightness temperature $\overline{T_{\rm HI}}=0.0050K$. The lognormal simulations of the continuous case produce $\overline{T_{\rm HI}}=0.0049\pm 0.0013K$ and in the discrete model produce $\overline{T_{\rm HI}}=0.0048\pm 0.001K$. The errors in these measurements increase with $\sigma_{\rm HI}$, the scatter with which the \HI\ masses per object were drawn from the \HI\ HOD.

\begin{figure}
\includegraphics[width=0.5\textwidth]{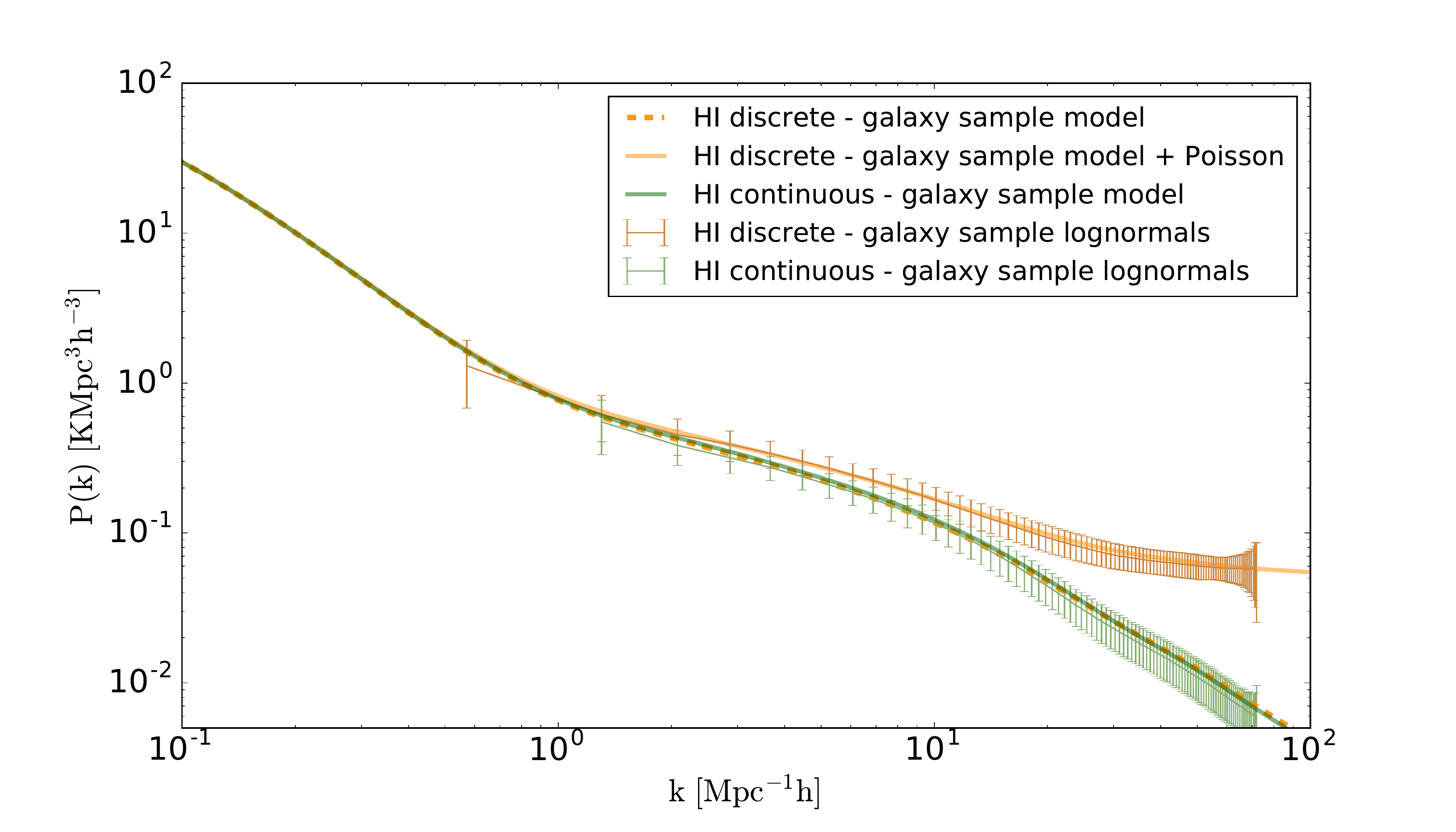}
\caption{The cross-power spectra predicted by \halomod\ for the \HI\ continuous model with the galaxy \textit{sample}, and the \HI\ discrete model with the galaxy \textit{sample}, in comparison with a average power spectrum of 100 lognormal realisations of a box of length $15 {\rm Mpc}/h$ drawn from the respective \HI\ and galaxy HOD. Note that the \HI\ continuous model does not include a scale-independent Poisson noise contribution since it is estimated from a continuum field. We show \halomod\ predictions including and excluding Poisson noise contribution.}
\label{FigPXHIcomp}
\end{figure}

The cross-power spectra of the galaxy sample with the two \HI\ models are presented in Fig.~\ref{FigPXHIcomp}. Even though the theory calculation of the two models does not predict any visible deviation on all considered scales, we observe that the inclusion of Poisson noise in the discrete model considerably increases the power in the range $k\gtrsim 2h/{\rm Mpc}$. The theoretical prediction of the cross Poisson noise is added to \halomod\ using Eq.~\ref{EqucrossSN}. As previously discussed, the cross Poisson noise scales with the \HI\ content of the galaxy population averaged over all halo masses, in this case for the galaxy \textit{sample}. For this galaxy \textit{sample}, we can measure an average \HI\ mass of $\log_{10}(\overline {M}_{\rm HI, sample}/M_\odot h)= 11.202 $ from the lognormal realisations, which is very close to the prediction of the theoretical model with $\log_{10}(\overline M_{\rm HI, sample}/M_\odot h)= 11.208 $. We note that in the considered lognormal realisations with volume $(15 {\rm Mpc}/h)^3$, the mean number of galaxies in the \textit{sample} is relatively low, with 121.

\subsection{Correlated Populations}
In the above example, the distribution of the \HI\ within the galaxy \textit{field} for each halo mass $m$ follows a Gaussian distribution with standard deviation $\sigma_{\rm HI}$. Thus there is no dependence of the \HI\ content on the galaxy occupation within the \textit{sample}. In reality, the amount of \HI\ present in the galaxy depends on its evolutionary state. In general terms, blue, star-forming galaxies are expected to be \HI -rich whereas red, quiescent galaxies are \HI -deficient. In this work, focusing on the concept of Poisson noise in intensity mapping, we do not concern ourselves with details such as luminosity functions which would be required to accurately model these dependencies.

\begin{figure}
\includegraphics[width=0.5\textwidth]{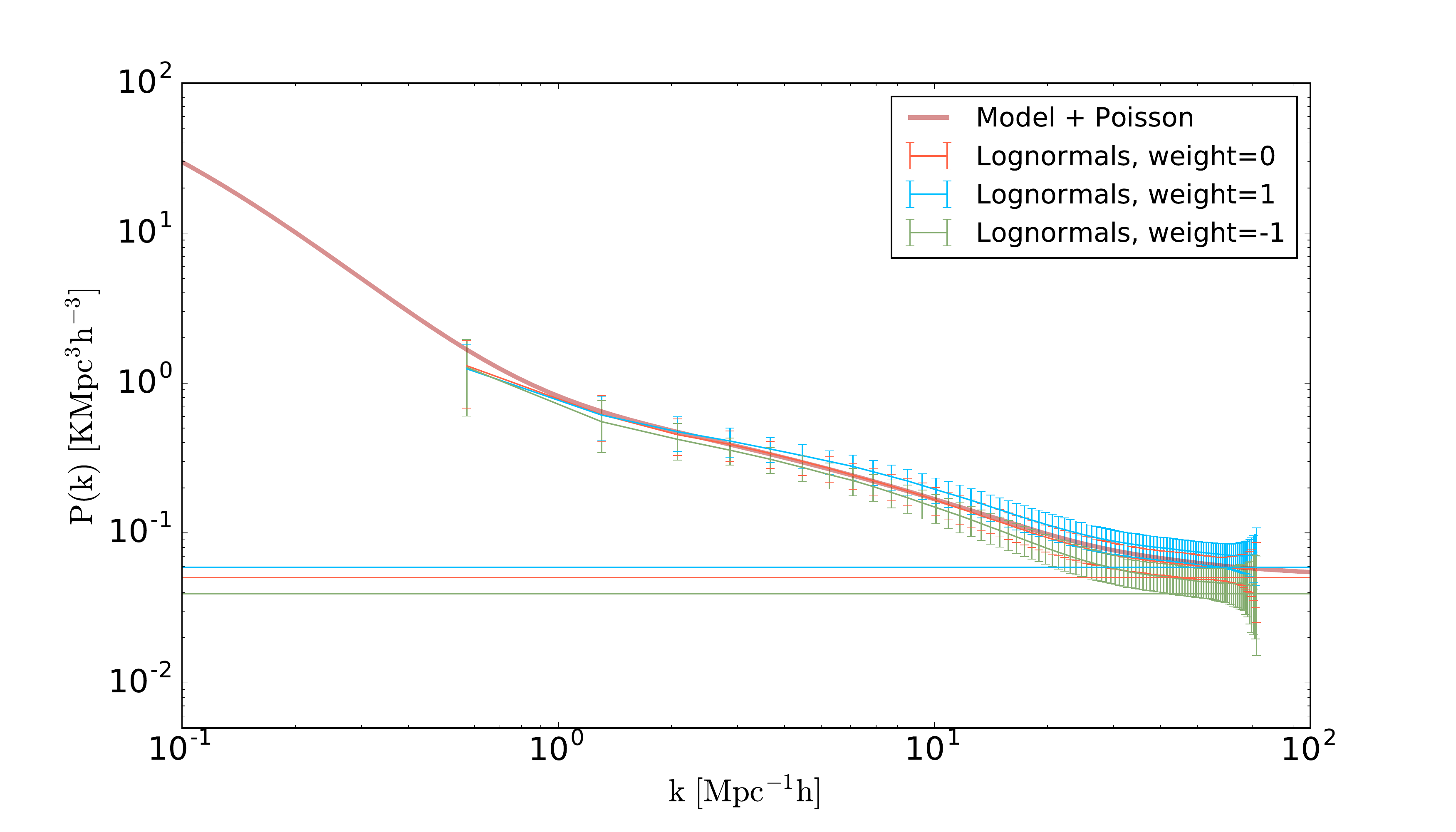}
\caption{The cross power spectra predicted by \halomod\ for the \HI\ discrete model with the galaxy \textit{sample} with different \HI\ weighting, in comparison with a average power spectrum of 100 lognormal realisations with a box of length $15 {\rm Mpc}/h$ drawn from the respective \HI\ and galaxy HOD. The dependence of the shape of the power spectra on the \HI\ weighting is negligible but the amplitude of the Poisson noise changes significantly. We show \halomod\ predictions including and excluding Poisson noise contribution.}
\label{FigPXHIcompweight}

\end{figure}

In order to mimic the effect that a correlation between luminosity and \HI\ mass would impose on the Poisson noise, we assume that the galaxy \textit{sample} describes the HOD of a specific type of galaxy which is correlated or anti-correlated with the \HI\ content as described in Sec.~\ref{SecHIpopcorr}. This correlation, as predicted, has no effect on the \HI\ auto-power or the shape of the cross-power, but it changes the amplitude of the cross-Poisson noise as the averaged \HI\ mass per galaxy in the \textit{sample} is modified. Fig.~\ref{FigPXHIcompweight} presents the result of the weighting of the \HI\ for the galaxy \textit{sample}. To demonstrate the change in the Poisson noise, we added the measured Poisson noise from the lognormal simulations with coloured horizontal lines to Fig.~\ref{FigPXHIcompweight}.

In general, a specific mathematical model of the correlation of two samples is not available, and so we cannot determine the Poisson noise amplitude \textit{a priori}. However, the \halomod\ theory predictions can be used as a tool to fit the pure Poisson noise contribution as well as measure the deviation compared to an uncorrelated sample.

\section{Parameter Estimation}
\label{Sec-MCMC}
In this section, we demonstrate the utility of the  \halomod\ algorithms to recover the parameters of a specific \HI\ model via a Monte-Carlo Markov Chain (MCMC) maximum likelihood fit. We use the Python package \textsc{emcee} \citep{2013PASP..125..306F} and fit the theory prediction of each \HI\ model to the estimated power spectra of  the lognormals with box size $(15 {\rm Mpc}/h)^3$, which optimally resolves the shot noise regime of the power spectra. We fit the averaged power spectra of 100 realisations, as well as 10 individual realisations. The covariance of each power spectrum measurement is given by the standard deviation of all realisations which includes Cosmic variance, fluctuations in the number densities due to the small box size and variations due to the population of \HI\ masses. We note that due to the limited size of the box, which measures a minimum scale $k\approx 1.0h/{\rm Mpc}$, our parameter fitting is limited to the scales dominated by the shot noise. The overall fitting could be improved using a wider range of wavenumbers, however, some intensity mapping experiments such as the interferometer ASKAP are only sensitive to similar scales $k> 1.0h/{\rm Mpc}$. For all cases, we only fit \HI\ parameters, efficiently employing very tight priors for parameters of the galaxy \textit{sample} distribution. We do not attempt to fit the maximum halo mass $M_{\rm max, HI}$ as for our tested box size, the abundance of these high mass halos is very low and the cut-off cannot be tested. 

\begin{figure}
\includegraphics[width=0.5\textwidth]{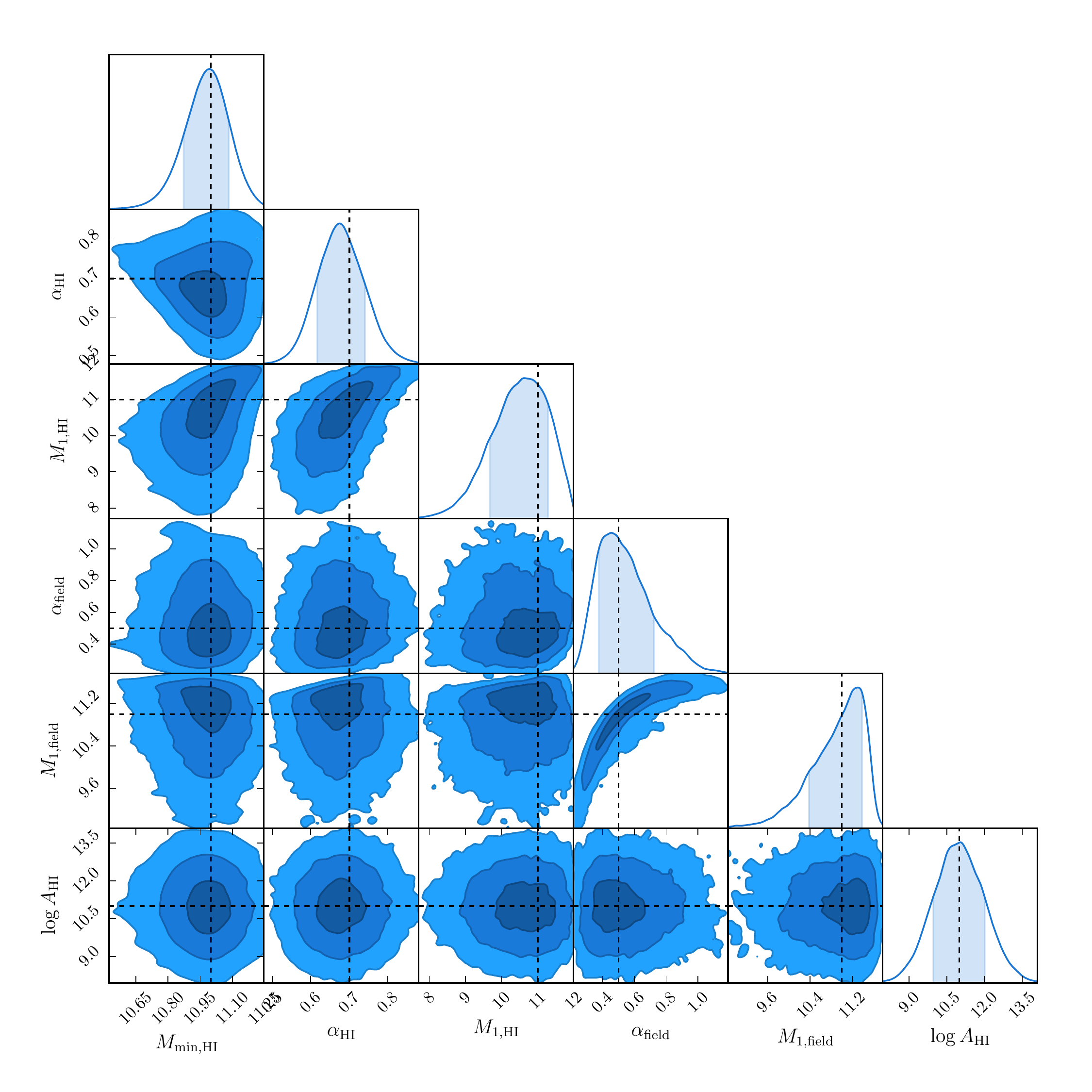}
\caption{The likelihood contours of the MCMC fit to the realisation-averaged power spectrum of the discrete \HI\ model, using the \HI\ auto-correlation and its Poisson noise to constrain the HOD parameters of \halomod. The dashed lines indicate the input parameter values. All masses are given as $\log_{10}$ and in units of $M_\odot/h$.}
\label{FigMCMCHIcs}
\end{figure}

We ran MCMCs with a total of $10^6$ samples and tested convergence of the chains via the Gelman-Rubin criteria where all parameters passed with a threshold of $R_{\rm GR}=1.1$. We set Gaussian priors with the standard deviations of all HOD parameters $\alpha$ given as $\sigma(\alpha_i)=0.3$ with $i=\{\rm HI, field\}$ and the standard deviation of all other HOD parameters as $\sigma(\theta_i)=1.0$. We note that our results, in particular for the discrete case in which the model is primarily fit to the constant amplitude of the Poisson noise, are not independent of the chosen priors. In particularly the \HI\ amplitude parameter, $\log{A_{\rm HI}}$, is constrained by the prior and can not be fit efficiently by the MCMCs unless a total temperature constraint is imposed. The best fit values are derived by cumulative statistics as the marginalised parameter likelihoods exhibit Non-Gaussian characteristics, as can be seen in Fig.~\ref{FigMCMCHIcs}.

\begin{figure}
\includegraphics[width=0.5\textwidth]{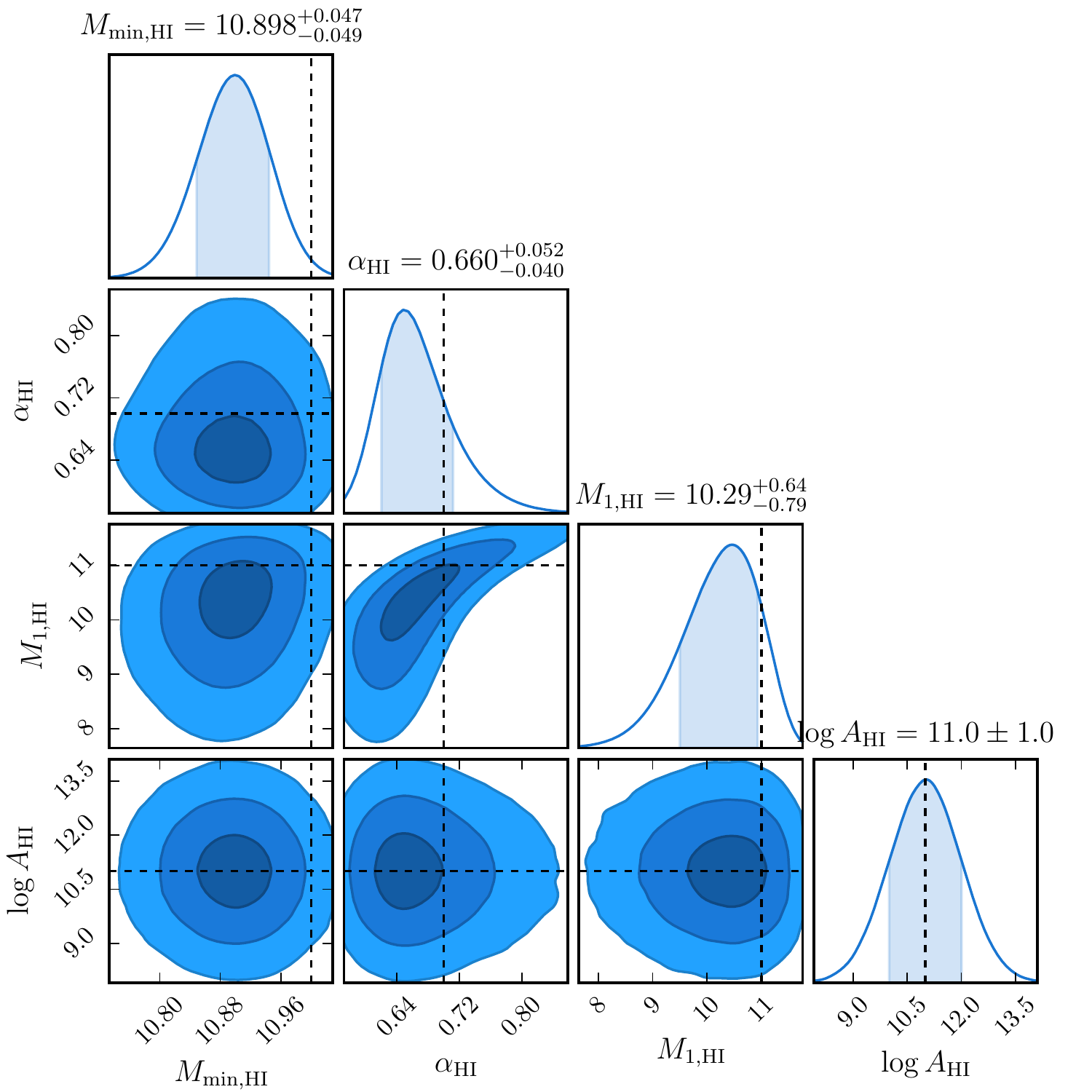}
\caption{The likelihood contours of the MCMC fit to the realisation-averaged power spectrum of the continuous \HI\ model, using the \HI-galaxy \textit{sample} auto-correlation and to constrain the HOD parameters of \halomod. The dashed lines indicate the input parameter values.  All masses are given as $\log_{10}$ and in units of $M_\odot/h$.}
\label{FigMCMCHIcont}
\end{figure}

The resulting likelihoods of the MCMCs of the auto-power spectra of both \HI\ models are displayed in Figs.~\ref{FigMCMCHIcs} and \ref{FigMCMCHIcont}.
In Tables~\ref{tab:HIcs}, and~\ref{tab:HIcont} we present the outcomes of the parameter estimation for the auto- and cross-power spectra of both \HI\ models. We individually fit the power spectra rather than perform a joint analysis as in many upcoming experiments only one or the other will be available due to limitations in the quality of data or lack of an optical galaxy sample. The parameter fits can be extremely biased due to the fluctuations in the lognormal realisations. In order to derive mean parameter fits and the expected variance including Cosmic variance while remaining computationally feasible, we run MCMCs on the mean power spectrum of all 100 realisations presented under names $\{ \rm{Auto}, ~\rm {Cross}\}$ in each table, in addition to running MCMCs on 10 individual realisations, and presenting the mean and standard deviation of their fits under $\{ \rm{Auto}\sum_{i=1}^{10}, ~\rm {Cross}\sum_{i=1}^{10} \}$  in the tables. Whereas the constraints given by the MCMCs of the mean demonstrate the degeneracy within the halo model parameters, the standard deviation over 10 realisation shows limitations due to Cosmic variance.  

Table~\ref{tab:HIcs} and Fig.~\ref{FigMCMCHIcs} presents the parameter constraints of the discrete \HI\ model, where the theory is primarily fit to the Poisson noise amplitude in the given $k$ range. From cross-correlation, we derive the ensemble-averaged \HI\ mass of the galaxy sample from the estimated parameters which is given by the numerator of the cross Poisson noise. The input parameters correspond to $\langle \log_{10}( M_{{\rm HI},g}/M_\odot h )\rangle=9.77$ and the mean of the $N=10$ realisations gives $\langle \log_{10}(M_{{\rm HI},g}/M_\odot h )\rangle_N=9.89\pm 0.33$. 

Table~\ref{tab:HIcont} and  Fig.~\ref{FigMCMCHIcont} display the results of the continuous \HI\ model, where only 4 parameters need to be estimated in the auto- and cross-correlation. The individual parameter constraints are much tighter due both to the fewer number of parameters, and the fact that the spectrum shape is not dominated by a single Poisson noise term. The uncertainty due to Cosmic variance are comparable to the \HI\ discrete case. 

\begin{table*}
    \centering
\caption{Discrete \HI\ model: Marginalised parameter likelihoods given by the MCMC fit to the averaged auto- and cross-power spectrum, demonstrating the degeneracies within the HOD parameters, and the mean and the standard deviation of the MCMC fit to 10 realisations marked with $\sum_{i=1}^{10}$, indicating the Cosmic variance. All masses are given as $\log_{10}$ and in units of $M_\odot/h$.}    \label{tab:HIcs}
    \begin{tabular}{ccccccc}
    
        \hline
		Model & $M_{\rm  min,HI}$ & $\alpha_{\rm HI}$ & $M_{1,\rm HI}$ & $\alpha_{\rm field}$ & $M_{1,\rm field}$ & $\log{A_{\rm HI}}$ \\ 
		\hline
		Input & 11.0  & 0.7 & 11.0 & 0.5 & 11.0 & 11.0  \\
		\hline \hline
		Auto  & $10.98^{+0.10}_{-0.11}$ & $0.677^{+0.064}_{-0.060}$ & $10.54^{+0.74}_{-0.88}$ & $0.52^{+0.20}_{-0.14}$ & $11.00^{+0.39}_{-0.62}$ & $11.0\pm 1.0$ \\ 
		\hline
		Auto $\sum_{i=1}^{10}$ & $ 10.945  \pm  0.2429 $ & $ 0.704  \pm  0.1388 $ & $ 10.67  \pm  0.522 $ & $ 0.548  \pm  0.0214 $ & $ 10.941  \pm  0.1453 $ & $ 11.007  \pm  0.0296 $\\
		\hline
		\hline
		Cross  & $10.97^{+0.18}_{-0.23}$ & $0.73\pm 0.12$ & $10.72^{+0.86}_{-0.95}$ & $0.55^{+0.22}_{-0.16}$ & $10.99^{+0.45}_{-0.66}$ & $11.01^{+1.00}_{-1.02}$ \\ 
		\hline
		Cross $\sum_{i=1}^{10}$ & $ 10.883  \pm  0.3171 $ & $ 0.753  \pm  0.1375 $ & $ 10.79  \pm  0.2711 $ & $ 0.544  \pm  0.0491 $ & $ 10.883  \pm  0.2265 $ & $ 10.998  \pm  0.0128 $
    \end{tabular}
\end{table*}
\begin{table*}
    \centering
\caption{Continuous \HI\ model: Marginalised parameter likelihoods given by the MCMC fit to the averaged auto- and cross-power spectrum, demonstrating the degeneracies within the HOD parameters, and the mean and the standard deviation of the MCMC fit to 10 realisations marked with $\sum_{i=1}^{10}$, indicating the Cosmic variance. All masses are given as $\log_{10}$ and in units of $M_\odot/h$.}
    \label{tab:HIcont}
    \begin{tabular}{ccccc}
        \hline
		Model & $M_{\rm  min,HI}$ & $\alpha_{\rm HI}$ & $M_{1,\rm HI}$ & $\log{A_{\rm HI}}$ \\ 
		\hline
		Input & 11.0 &  0.7 & 11.0 & 11.0  \\
		\hline
		\hline
		Auto & $10.898^{+0.047}_{-0.049}$ & $0.660^{+0.052}_{-0.040}$ & $10.29^{+0.64}_{-0.79}$ & $11.0\pm 1.0$ \\ 
		\hline
		Auto $\sum_{i=1}^{10}$ &	$ 10.852  \pm  0.3594 $ & $ 0.611  \pm  0.1013 $ & $ 10.347  \pm  0.4343 $ & $ 11.004  \pm  0.0087 $ \\
		\hline
		\hline
		Cross & $10.866^{+0.106}_{-0.095}$ & $0.714^{+0.083}_{-0.075}$ & $10.61^{+0.84}_{-0.90}$ & $11.0\pm 1.0$ \\ 
		\hline
        	Cross $\sum_{i=1}^{10}$ & $ 10.877  \pm  0.3458 $ & $ 0.626  \pm  0.1561 $ & $ 10.816  \pm  0.3106 $ & $ 10.998  \pm  0.0093 $\\
		
    \end{tabular}
\end{table*}

\section{Summary and Discussion}
\label{Sec-sum}
In this study we present a new, adaptive description of the intensity mapping auto-power spectrum and cross-power spectrum with galaxy surveys in the halo model framework (using \halomod). We introduce two different implementations for the description of \HI\ populations; the continuous \HI\ model which populates \HI\ within Dark Matter halos following a smooth profile, and the discrete \HI\ model which co-locates \HI\ masses with the positions of an underlying galaxy field, where \HI\ and field can follow independent HOD descriptions. The models represent the opposite ends of the spectrum of currently used \HI\ simulations. We inspect the impact of the different \HI\ models on the shapes of the auto- and cross-power spectra and find that the \HI\ power spectra of both models only differ on scales $k > 10 h/{\rm Mpc}$, caused by the additional central-satellite contributions in the 1-halo term of the discrete model. The prediction of the 1-halo terms of both \HI\ models in the cross-power spectrum with galaxies are very similar if the same \HI\ halo profiles are used. 

We verified our analytic predictions with a set of lognormal realisations, and find that the major difference between the models is the presence or absence of shot noise contributions. We review the current understanding of shot noise in galaxy and \HI\ intensity mapping data and state analytic expressions to determine the amplitude of the shot noise given the underlying HODs. Most notably, the shot noise on the cross-power spectrum directly scales with the averaged \HI\ mass of the optical galaxies, which is well-defined in the halo model framework.

We examine the shot noise properties of both \HI\ models and find that the implementation of the continuous \HI\ models has no Poisson noise contribution to any power spectra due to the continuous, smooth nature of the \HI\ density field. The shot noise of the discrete \HI\ model is correctly predicted by \halomod\ for the auto-correlation and cross-correlation with a galaxy \textit{sample}, given the \HI\ content is independent of the galaxy sample abundances. In our examples, the Poisson noise contributions dominate the amplitude of the overall power spectra from scales $k>10h/{\rm Mpc}$.The cross-Poisson noise is proportional to the averaged \HI\ mass per galaxy in the \textit{sample} and, as such, can be used to determine the average \HI\ mass of galaxy samples without directly observing their \HI\ content. Our \halomod\ implementation is the first tool to predict the cross-Poisson noise given \HI\ and galaxy HOD parameters and will be useful for experimental forecasts as well as observational interpretations.  

We demonstrate how the \HI\ model parameters of the \halomod\ predictions can be fit to the simulations using MCMC techniques. These fits also estimate derived \HI\ properties such as the average brightness temperature, which is directly proportional to $\Omega_{\rm HI}$, and the averaged \HI\ mass per galaxy in the cross-correlation with galaxy \textit{sample}, a quantity of great interest in future cross-correlation experiments on small scales. This way, \halomod\ has the potential to estimate the unknown parameters of the \HI\ distribution traced by the \HI\ intensity maps, as well as determining the averaged \HI\ masses of galaxy samples in intensity mapping cross-correlation experiments.

We note that our study exclusively focuses on the impact of the halo occupation parameters on the power spectra and Poisson noise. We have not considered the degeneracy of cosmological parameters and halo occupation parameters, but on the scales considered in this work the effect of galaxy evolution dominates. We note that non-linear effects of the power spectrum and peculiar velocities alter the shape and amplitude of the 1-halo term, however, on small enough scales, the contribution of the Poisson noise is considerably more dominant than the 1-halo term. In these cases, the \halomod\ prediction of the cross-Poisson noise could be added to a more sophisticated power spectrum model which includes these effects or the fits could be performed to the projected correlation function to suppress redshift-space distortions.

In this project, we did not employ data-motivated \HI\ models as we aim to demonstrate a maximally flexible framework for \HI\ auto- and cross-power spectrum and their Poisson noise predictions that can be adapted to individual experiments' needs. Our model allows us to easily import any shape and parametrisation of the \HI\-to-halo relation and examine their predictions. In future work, more data-driven models will be implemented in order to compare predictions with observations.

\section*{Acknowledgements}
LW is supported by an ARC Discovery Early Career Researcher Award (DE170100356).
This research was conducted by the Australian Research Council Centre of Excellence for All-sky Astrophysics (CAASTRO), through project number CE110001020. This work was performed on the gSTAR national facility at Swinburne University of Technology. gSTAR is funded by Swinburne and the Australian Government's Education Investment Fund.




\bibliographystyle{mnras}
\bibliography{bib} 

\appendix
\section{Correlation models for HI-galaxy samples}
\label{Appcorr}
\subsection{Continuous HI Model}
\label{Appcorrcont}
In this model, the gas is not co-located with observed galaxies, forming a spatially-independent smooth profile within the halo. 
In reality, we expect that while the \textit{averaged} profile of the gas is smooth, it will be lumpy on scales much smaller than the halo radius. 
It is then expected that the prospects of observing a galaxy in the sample may be dependent on the \HI\ density around the location of the galaxy.
However, dealing with this general situation, in which spatial scales within the halo are correlated according to the typical size of the \HI\ ``lumps'' is rather difficult, and we may consider two extreme cases in more detail. 
The first is that in which the lumps are infinitely broad, or rather that the \HI\ profile is perfectly smooth for every halo.
The second is that in which the ``lumps'' are Dirac-$\delta$ functions, but this is equivalent to the discrete \HI\ model which we consider in the following subsection.

Suppose that the \HI\ profile of every halo is always completely smooth, and is constant with the underlying halo mass. 
Suppose also that there is a distribution of \HI\ masses for a given halo mass, for which the mean is $\langle M_{\rm HI}(m) \rangle$, and the variance is $\sigma_{\rm HI}^2$ (the distribution remains arbitrary, but one may like to think of it as a Gaussian or Lognormal). 
If a particular halo has an \HI\ mass from the upper-tail of its distribution, then the \HI\ density of that halo is increased uniformly everywhere in the halo, because it is necessarily completely smooth.
Now consider a sample of observed galaxies. 
The probability of finding a galaxy at any point in a given halo may depend on the density of the \HI\ in that location (in fact, it may depend on much more than that, for example, it may depend on the \HI\ density in nearby locations, or the dynamical state of the \HI\ rather than just its abundance, but these are considered to be minor complications which we will ignore). 
However, since the density of \HI\ at any given location is determined by the density at all other locations, or rather, the density at any location is fully specified by the total \HI\ mass in the halo -- due to its smoothness -- the total expected number of observed galaxies in the halo is completely determined by its \HI\ mass. 
Summarily, we have the following system:
\begin{align}
    M^i_{\rm HI} &\sim \phi(m_{\rm HI},m), \nonumber \\
\   N^i &\sim {\rm Poisson}(f(M^i_{\rm HI})),
\end{align}
where $f$ is some function which converts the actual \HI\ mass of a halo into the expected number of observed galaxies. 
While this function is arbitrary, a simple but flexible toy model is such that $f = n_0 (M^i_{\rm HI} / A)^{\gamma}$.
Letting $\tilde{A}(m)$ be the average amount of \HI\ per galaxy in halos of mass $m$, we obtain that $\langle N \rangle = \langle M_{\rm HI} \rangle/\tilde{A}$.
Furthermore we find that
\begin{equation}
    \tilde{A}  = \frac{A^\gamma}{n_0} \frac{\langle M_{\rm HI} \rangle}{\langle M_{\rm HI}^\gamma \rangle}.
\end{equation}
We focus hereafter on some special cases, $\gamma \in (-1,0,1)$, corresponding to anti-correlated, uncorrelated and correlated cases. 
For these we have
\begin{equation}
    \tilde{A} = \begin{cases} (A n_0)^{-1} \frac{\langle M_{\rm HI} \rangle}{\langle M^{-1}_{\rm HI} \rangle}, &\gamma = -1 \\
    \frac{\langle M_{\rm HI} \rangle}{n_0}, &\gamma = 0 \\
    \frac{A}{n_0}, &\gamma =1.
	\end{cases}
\end{equation}
We note that the distribution of $N$ in such a setup is not necessarily Poisson, as we generally assume it to be.
Nevertheless, it is not likely to be significantly different to Poisson, and in any case, this fact does not affect the rest of our calculations.

We wish to calculate the value of $\langle N_s M_{\rm HI} \rangle$.
This can be achieved by using the law of total expectation,
\begin{align}
  \langle N_s M_{\rm HI} \rangle  &= \int dm' m' \phi(m',m) \sum_{k=0}^\infty k {\rm Pois}(f(m')) \nonumber \\
  &= \sum_{k=0}^\infty \frac{n_0}{A^{\gamma k} (k-1)!} \int dm' \ m'^{\gamma k+1} \phi(m',m) e^{-n_0 (m'/A)^\gamma}.
\end{align}

If we assume that $\phi$ is a Gaussian distribution, then for masses $m$ at which the expected number of galaxies is large, since the Poisson distribution tends to a Gaussian, the result tends to
\begin{equation}
    \langle N_s M_{\rm HI} \rangle = \frac{n_0 A^{-\gamma}}{\sqrt{2\pi}\sigma} \int dm' \ m'^{1+\gamma} e^{-(m' - \langle M_{\rm HI} \rangle)^2/2\sigma_{\rm HI}^2}.
\end{equation}
While this is in general unsolvable, it yields solutions for our three cases of interest:
\begin{equation}
    \langle N_s M_{\rm HI} \rangle \approx \begin{cases} n_0 A , &\gamma = -1 \\
    n_0 \langle M_{\rm HI} \rangle = \langle N \rangle \langle M_{\rm HI} \rangle, &\gamma = 0 \\
    n_0 \frac{\langle M_{\rm HI} \rangle^2 + \sigma_{\rm HI}^2}{A} , &\gamma = 1. \end{cases}
\end{equation}

Thus the correlation function is (in the large-$N$ limit):
\begin{align}
    R(m) &= \frac{\langle N_s M_{\rm HI} \rangle - \langle N \rangle \langle M_{\rm HI} \rangle}{ \sqrt{\langle N \rangle} \sigma_{\rm HI}}  \nonumber \\
    &= \begin{cases}
    \frac{1}{\sigma_{\rm HI}}\sqrt{\frac{\langle M_{\rm HI}\rangle}{\tilde A}}\left( \frac{1}{ \langle M^{-1}_{\rm HI} \rangle} -\langle M_{\rm HI} \rangle\right), &\gamma = -1 \\
    0, &\gamma = 0 \\
   \frac{\sigma_{\rm HI}}{\sqrt{\tilde{A}\langle M_{\rm HI} \rangle}}, &\gamma = 1.
\end{cases}
\end{align}

The question of how to calculate $\langle M^{-1}_{\rm HI} \rangle$ remains.
We find that a useful empirical formula is such that
\begin{equation}
  \langle M^{-1}_{\rm HI} \rangle \approx \frac{\langle M_{\rm HI} \rangle^2 + \sigma_{\rm HI}^2}{\langle M_{\rm HI} \rangle^3}
\end{equation}
when $M_{\rm HI}$ is a Gaussian variable, and $\langle M_{\rm HI} \rangle/\sigma_{\rm HI} > 3$.
This latter condition must be obeyed in any case to ensure the description is physically appropriate, otherwise a significant part of the probability density puts $M_{\rm HI} < 0$.
Under this approximation, the correlation function becomes
\begin{equation}
  R_{\gamma = -1}(m) = - \sigma_{\rm HI} \langle M_{\rm HI} \rangle \frac{\sqrt{\frac{M_{\rm HI}}{\tilde{A}}}}{ \langle M_{\rm HI} \rangle^2 + \sigma_{\rm HI}^2}
\end{equation}

\subsection{Discrete HI Model}
\label{Appcorrdisc}
The result in the case of the discrete model, in which the ``lumps'' are Dirac-$\delta$ functions, is much simpler.
The typical assumption of the galaxies obeying a Poisson distribution carries with it the assumption that they are spatially independent, and we have by construction specified that the \HI\ components are also spatially independent.
This implies that while the probability of observation of a galaxy at a certain point may be dependent on the \HI\ in that location, it is entirely uncorrelated with any other point.
This means that all correlations exist at a separation of zero, which is not represented in the power spectrum at all.
Alternatively one may consider Eq.~\ref{EquPXcorr}, in which the total contribution of the satellite-satellite term has a $-Q$ term, which accounts for all self-pairs.
After subtracting the self pairs, no other pairs contain any correlations, and so in general we have that
\begin{equation}
  \langle N_s M_{\rm HI} \rangle  - Q = \langle N_s \rangle \langle M_{\rm HI} \rangle.
\end{equation}

Nevertheless, while these correlations cannot change the \textit{shape} of the power spectrum, they do affect the level of shot-noise present.
This is simple to conceptualise; since the shot-noise depends on the average \HI\ mass within galaxies of the \textit{sample}, a correlation which favours observing galaxies which contain a higher \HI\ mass will therefore accordingly increase the shot-noise, and vice versa.
The net result is a constant additive factor to the observed power spectrum, and thus detailed modelling will not usually be required --- the constant may just as well be fit and then interpreted.

In practice, one may conceive of the correlation occurring in multiple ways.
In any of these, if the mean \HI\ mass of the sample can be calculated, it can be used to directly infer the amplitude of the shot noise.


\bsp	
\label{lastpage}
\end{document}